\documentclass[sigconf]{acmart}
\AtBeginDocument{%
  }

\setcopyright{acmlicensed}

\copyrightyear{2026}
\acmYear{2026}
\setcopyright{cc}
\setcctype{by}
\acmConference[KDD 2026] {Proceedings of the 32nd ACM SIGKDD Conference on Knowledge Discovery and Data Mining V.1}{August 9--13, 2025}{Jeju Island, Republic of Korea.}
\acmBooktitle{Proceedings of the 32nd ACM SIGKDD Conference on Knowledge Discovery and Data Mining V.1 (KDD 2026), August 9--13, 2025, Jeju Island, Republic of Korea}
\acmISBN{979-8-4007-2258-5/2026/08} 
\acmDOI{10.1145/3770854.3780314}

\usepackage{graphicx}
\usepackage{amsmath}
\usepackage{bm}
\usepackage{amsthm}
\theoremstyle{plain}
\newtheorem{theorem}{Theorem}[section]
\newtheorem{proposition}[theorem]{Proposition}
\usepackage{float}
\usepackage{colortbl}
\usepackage{wrapfig}
\usepackage{multirow}
\usepackage{subfigure}
\usepackage{algorithm}
\usepackage{algorithmic}

\usepackage{amssymb}
\newtheorem{assumption}{Assumption}[section]

\definecolor{CadetBlue}{rgb}{0.37, 0.62, 0.63}
\definecolor{Cerulean}{RGB}{0,123,167}     
\definecolor{BurntOrange}{RGB}{204,85,0}   

\begin{document}

\title{Self-Guided Diffusion Model for Accelerating Computational Fluid Dynamics}

\author{Ruoyan Li}
\orcid{0009-0009-5098-7900}
\authornote{Equal Contribution.}
\email{liruoyan2002@g.ucla.edu}
\affiliation{%
  \institution{University of California, Los Angeles}
  \city{Los Angeles}
  \state{California}
  \country{USA}
}

\author{Zijie Huang}
\orcid{0009-0008-1263-9270}
\authornotemark[1]
\email{zijiehuang@cs.ucla.edu}
\affiliation{%
  \institution{University of California, Los Angeles}
  \city{Los Angeles}
  \state{California}
  \country{USA}
}

\author{Haixin Wang}
\orcid{0000-0002-5714-0149}
\affiliation{%
  \institution{University of California, Los Angeles}
  \city{Los Angeles}
  \state{California}
  \country{USA}
}
\email{whx@ucla.edu}

\author{Guancheng Wan}
\orcid{0000-0002-7083-6423}
\affiliation{%
  \institution{University of California, Los Angeles}
  \city{Los Angeles}
  \state{California}
  \country{USA}
}
\email{gcwan03@ucla.edu}

\author{Yizhou Sun}
\orcid{0000-0003-1812-6843}
\affiliation{%
 \institution{University of California, Los Angeles}
  \city{Los Angeles}
  \state{California}
  \country{USA}
}
\email{yzsun@cs.ucla.edu}

\author{Wei Wang}
\orcid{0000-0002-8180-2886}
\affiliation{%
  \institution{University of California, Los Angeles}
  \city{Los Angeles}
  \state{California}
  \country{USA}
}
\email{weiwang@cs.ucla.edu}

\renewcommand{\shortauthors}{Li et al.}

\newcommand{\model}{SG-Diff}
\begin{abstract}
  Machine learning methods, such as diffusion models, are widely explored as a promising way to accelerate high-fidelity fluid dynamics computation via a super-resolution process from faster-to-compute low-fidelity input. However, existing approaches usually make impractical assumptions that the low-fidelity data is down-sampled from high-fidelity data. In reality, low-fidelity data is produced by numerical solvers that use a coarser resolution. Solver-generated low-fidelity data usually sacrifices fine-grained details, such as small-scale vortices compared to high-fidelity ones. Our findings show that SOTA diffusion models struggle to reconstruct solver-generated low-fidelity inputs. To bridge this gap, we propose SG-Diff, a novel diffusion model for reconstruction, where both low-fidelity inputs and high-fidelity targets are generated from numerical solvers. We propose an \textit{Importance Weight} strategy during training that serves as a form of self-guidance, focusing on intricate fluid details, and a \textit{Predictor-Corrector-Advancer} SDE solver that embeds physical guidance into the diffusion sampling process. Together, these techniques steer the diffusion model toward more accurate reconstructions. Experimental results on four 2D turbulent flow datasets demonstrate the efficacy of \model~against state-of-the-art baselines.
  Code and datasets are available at \href{https://github.com/RuoyanLi2002/Self-Guided-Diffusion-Model-for-Accelerating-Computational-Fluid-Dynamics.git}{Github}.
\end{abstract}




\maketitle

\section{Introduction}
High-fidelity simulations of computational fluid dynamics (CFD) are crucial for understanding fluid interactions in engineering systems, greatly impacting design and application outcomes~\citep{wang2024recent, wang2025fdbenchmodularfairbenchmark}. Traditional approaches such as Direct Numerical Simulation (DNS) offer high-resolution solutions. However, they are computationally expensive, especially for turbulence with high Reynolds numbers~\citep{zhang2023review}. Therefore, learning neural-based simulators from data becomes an attractive alternative, balancing between efficiency and simulation fidelity~\citep{huang2023generalizing, li2025flow}. 

\begin{table*}[t]
\centering
\small
\begin{tabular}{|
  >{\centering\arraybackslash}p{0.12\textwidth}
  |>{\centering\arraybackslash}p{0.16\textwidth}
  |>{\centering\arraybackslash}p{0.16\textwidth}
  |>{\centering\arraybackslash}p{0.16\textwidth}
  |>{\centering\arraybackslash}p{0.16\textwidth}
  |>{\centering\arraybackslash}p{0.12\textwidth}
|}
\hline
\textbf{Model Type} & \textbf{Training Data} & \textbf{Learning Goal} & \textbf{Advantages} & \textbf{Drawbacks} & \textbf{Examples} \\ 
\hline
\textcolor{BurntOrange}{Direct Mapping} & Paired low‑ and high‑fidelity data & Learn a specific mapping from low to high fidelity data & Fast inference & Must retrain if new low‑fidelity resolution differs from training & CNNs \\ 
\hline
\textcolor{Cerulean}{Distribution Learning} & High‑fidelity data only & Capture the probability distribution of high fidelity data & Can handle low‑fidelity inputs at arbitrary resolution & Slower to generate samples & Diffusion models \\ 
\hline
\end{tabular}
\caption{Comparison of Direct Mapping and Distribution Learning Approaches}
\label{tab:method_catg}
\end{table*}

One popular strategy is to reconstruct high-fidelity data from low-fidelity inputs, where the input data usually reduces the discretization grid size in the spatial domain and thus is  computational-efficient~\citep{shu2023physics, pradhan2021variationalmultiscalesuperresolution}. Various machine learning models, including those based on Convolutional Neural Networks (CNNs) \citep{Fukami_Fukagata_Taira_2019}, Generative Adversarial Networks (GANs) \citep{Generative}, and Diffusion Models \citep{shu2023physics, shan2024pirdphysicsinformedresidualdiffusion}, have been developed to reconstruct high-fidelity CFD data from low-fidelity inputs. The majority of them are categorized as \textcolor{BurntOrange}{direct mapping models}, which require both low- and high-fidelity data for training and can only capture a particular resolution mapping. In contrast, \textcolor{Cerulean}{distribution learning approaches}, such as diffusion models \citep{shu2023physics}, only require high-fidelity data during training and learns the probability distribution of high-fidelity sources. Therefore, they can reconstruct from low-fidelity data at any resolution without retraining and are particularly useful when large-scale solver-generated low-fidelity data are unavailable. We compare these two approaches in Table~\ref{tab:method_catg}.

\begin{figure}[t]
  \centering
  \includegraphics[width=0.5\textwidth]{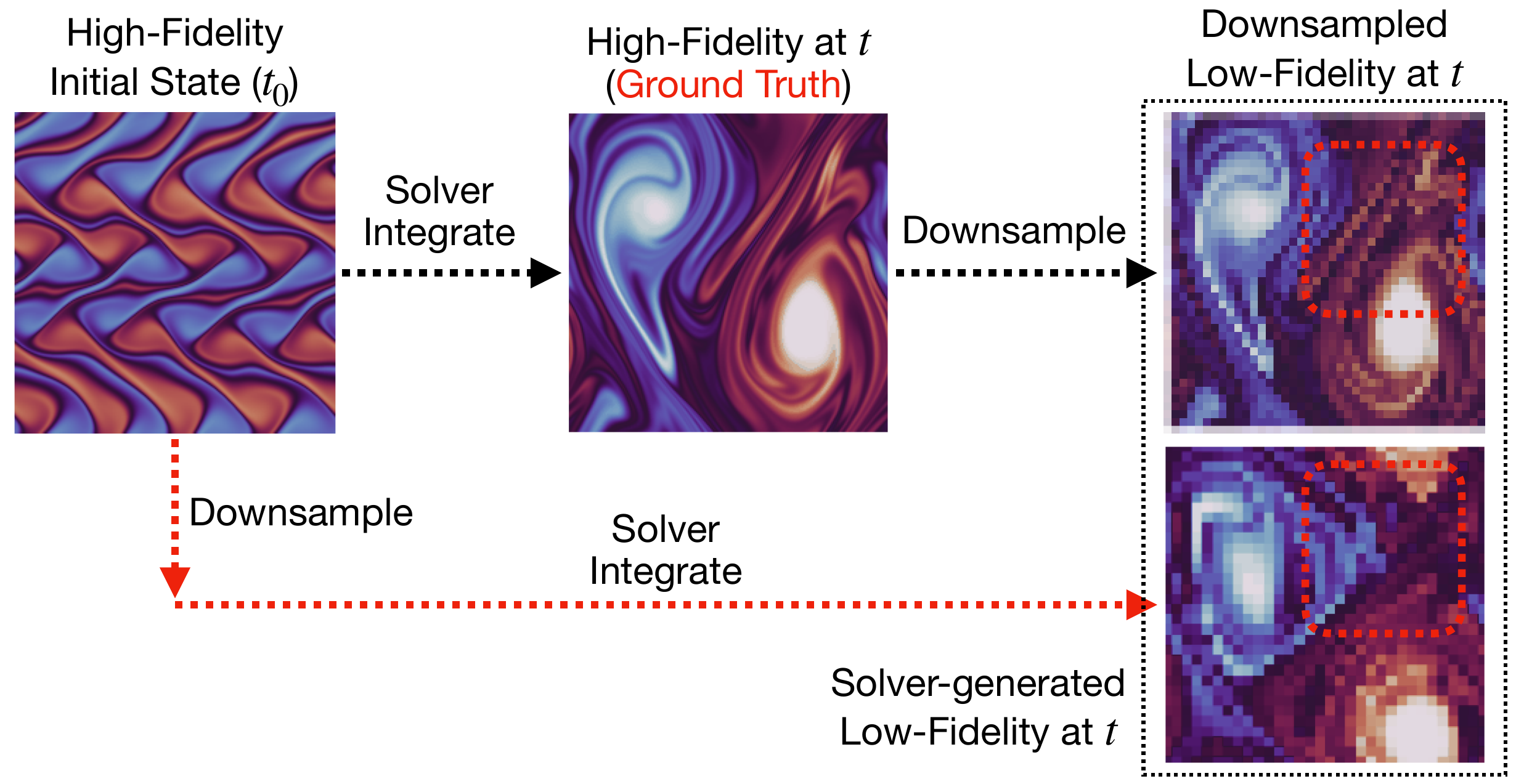}
  \caption{Comparison between downsampled (\textbf{black} line) and solver-generated (\textcolor{red}{\textbf{red}} line) flow fields. Solver-generated low-fidelity data retain less information, especially for fine-grained high-fidelity details.}
  \label{fig:motivation}
\end{figure}

One fundamental drawback in existing models is that low-fidelity data is artificially downsampled from high-fidelity data at the same timestamp. Such data inherently has more information compared to solver-generated low-fidelity data in reality, where coarser discretization grids are used in numerical solvers to save computational resources. As illustrated in Figure~\ref{fig:motivation}, the former follows ``integrate then downsample", which starts from the high-fidelity initial states to rollout trajectories, and then downsample at each timestamp. The latter follows ``downsample then integrate", which downsamples the high-fidelity initial state to obtain coarser discretization grids as starting points and rollout trajectories through numerical solvers.

Our experiments reveal that SOTA diffusion models struggle with solver-generated low-fidelity inputs, and we show why they fail. As a result, \textbf{we study how to use \textcolor{Cerulean}{distribution learning approaches}, specifically, diffusion models, for reconstructing high-fidelity CFD data from solver-generated low-fidelity data}. In response to this challenge, we build upon the state-of-the-art model proposed by \citet{shu2023physics} and propose an \textit{Importance Weight} strategy during training as self-guidance to locate fine-grained high-fidelity details, and a \textit{Predictor-Corrector-Advancer} stochastic differential equation (SDE) solver during inference for ensuring physical consistency. The two modules jointly guide the diffusion model toward higher-quality reconstructions.

Our key contributions are summarized as follows: 
\begin{itemize}
    \item \textbf{Problem Identification.} We study a novel problem on reconstructing high-fidelity flow fields with solver-generated low-fidelity data for distribution learning models. Our evidence reveals that SOTA diffusion models fail to generate high-quality outputs.
    \item \textbf{Practical Solution.} In light of this issue, we propose \model, a novel diffusion model with \textit{Importance Weight} strategy during training as self-guidance and a \textit{Predictor-Corrector-Advancer} SDE solver for physically coherent sampling.
    \item \textbf{Experimental Validation.} We present empirical evidence of \model’s superior performance in a variety of
    2D turbulent flow over 4 datasets. It yields a significant improvement in terms of predictive accuracy, physical consistency, and perceptual quality. 
\end{itemize}

\section{Preliminaries and Related Work}
We consider a machine learning model $f_{\theta}: \mathcal{X} \rightarrow \mathcal{Y}$ with parameters $\theta$, which transforms a data sample from low-fidelity domain $x\in \mathcal{X} \subseteq \mathbb{R}^{m \times m}$ to high-fidelity domain $y \in \mathcal{Y} \subseteq \mathbb{R}^{n \times n} (m<n)$. The distributions of the training and test sets for low-fidelity data are denoted by $p_{\mathcal{X}}^{\text{train}}$ and $p_{\mathcal{X}}^{\text{test}}$, respectively, and for high-fidelity data as $p_{\mathcal{Y}}^{\text{train}}$ and $p_{\mathcal{Y}}^{\text{test}}$. The objective is to develop $f_{\theta}$ such that it can effectively map samples from $\mathcal{X}^{\text{test}}$ to their corresponding high-fidelity counterparts $\mathcal{Y}^{\text{test}}$\footnote{For direct mapping models, the training takes low- and high-fidelity pairs as inputs. For diffusion models such as \citet{shu2023physics}, the training only requires high-fidelity data, and low-fidelity data is used as input during testing.}. As diffusion model operates on the same input and output grid, we upsample the low-fidelity data uniformly during inference as the model input.

\subsection{AI for Computational Fluid Dynamics (CFD)}
\textbf{Flow Field Reconstruction} In the field of high-fidelity CFD reconstruction, researchers have developed powerful models rooted in image super-resolution domain in computer vision. For example, \citet{PhySR,Jiang2020MESHFREEFLOWNETAP} leveraged CNN for spatial-temporal super-resolution. 
\citet{Generative} further advanced the field by proposing physics-informed GANs. Additionally, \citet{fu2023semisupervised} proposed a refinement network to address the issue with limited high-fidelity data. These models rely on low- and high-fidelity data pairs during training and can only capture specific resolution mappings. As a result, \citet{shu2023physics} leveraged diffusion model, which is trained exclusively on high-fidelity data and enables reconstruction from any low-fidelity input at any resolution. However, their experiments assume the low-fidelity data is artificially downsampled from high-fidelity sources, which leads to unreliable results. \citet{sarkar2023redefiningsuperresolutionfinemeshpde} evaluate direct mapping models that are trained and tested on solver-generated low-fidelity data. These approaches depend on large collections of low- and high-fidelity pairs from numerical solvers for training, and degrade once the low-fidelity inputs deviate from their training distribution. 

\textbf{PDE Simulation}
Recent advances in AI for PDE simulation include neural operator approaches like DeepONet \citep{lu2019deeponet} and Fourier Neural Operators \citep{li2020fourier}, which learn mappings between infinite-dimensional function spaces to generalize across discretizations. Extensions such as Physics-Informed Neural Operators \citep{li2021physics} incorporate PDE structure into training. Latent generative models have also emerged: Text2PDE \citep{voleti2024text2pde} applies diffusion models conditioned on text prompts, while flow-matching accelerates high-resolution forecasting \citep{wildberger2023flow}. MultiPDENet \citep{song2024multipdenet} integrates Runge-Kutta schemes with learnable convolutions for long-term accuracy on fluid dynamics tasks.

\subsection{Diffusion Model}\label{sec:prelim_DDPM}
Diffusion models have become a prominent class of deep generative models, demonstrating state-of-the-art performance across various domains such as image generation \citep{meng2022sdedit, saharia2021image, ho2020denoising, li2025deepgenerativemodelshard}, video synthesis \citep{inproceedings, yu2022digan}, and applications in scientific fields \citep{shu2023physics, yang2023denoisingdiffusionmodelfluid, qiu2024pifusionphysicsinformeddiffusionmodel}. Diffusion models are grounded in a stochastic diffusion process, akin to those found in thermodynamics. It contains a forward and reverse process, where a data sample is gradually corrupted with noise, and a neural network is trained to reverse this. The forward process is defined by the SDE:
\begin{align}
    dx = f(x,t) dt + g(t) dW_t,\ x_0 \sim p_{\text{data}}(x),
\end{align}
where $f(x,t)$ is the drift function, $g(t)$ is a time-dependent diffusion coefficient, and $dW_t$ is Brownian motion. $t \in \left[ 0, T \right]$ is a continuous time variable with $t=0$ corresponding to original data and $t=T$ corresponding to pure noise. The reverse diffusion is given by
\begin{align} \label{eqn:reverse_sde}
    dx = \left[ f(x,t) - g(t)^2 \nabla_x \log p_t (x) \right] dt + g(t) d\overline{W}_t,
\end{align}
where $\nabla_x \log p_t (x)$ is the score function of the intractable marginal distribution and $d\overline{W}_t$ is the reverse-time Brownian motion. We leverage a parametrized score model $s_{\theta} (x_t, t)$ to estimate the score function and use the score matching objective \citep{song2021scorebased} for training:
\begin{align} \label{eqn:loss_DDPM_original}
    \mathbb{E}_{t, x_0, x_t} \left[ \lambda(t) \| s_{\theta} (x_t, t) - \nabla_{x_t} \log p_t (x_t \mid x_0 ) \|^2 \right],
\end{align}
where $t \sim \text{Uniform}(0, T)$ and $x_t \sim p(x_t \mid x_0)$. 

\textbf{Diffusion Model for Linear Inverse Problem} Diffusion models have become a powerful class of priors for solving linear inverse problems such as image super-resolution. Early works like DiffPIR \cite{zhu2023denoising} and ILVR \cite{choi2021ilvr} introduced plug-and-play frameworks where a pre-trained unconditional diffusion model is conditioned post hoc. Building on this, DPS \cite{chung2023diffusion} leverages a posterior sampling formulation to guide the reverse process using a likelihood score. More efficient variants have emerged, including ResShift \cite{yue2023resshiftefficientdiffusionmodel}, which modifies the diffusion trajectory to accelerate sampling, and RePaint \cite{lugmayr2022repaint}, which introduces re-noising steps for temporal consistency. In super-resolution specifically, methods like SR3 \cite{saharia2022sr3} and DDNM \cite{wang2022ddnm} push toward task-specific designs or refinement strategies to enhance quality or controllability.

\begin{figure*}[t]
    \centering
    \includegraphics[width=0.8\linewidth]{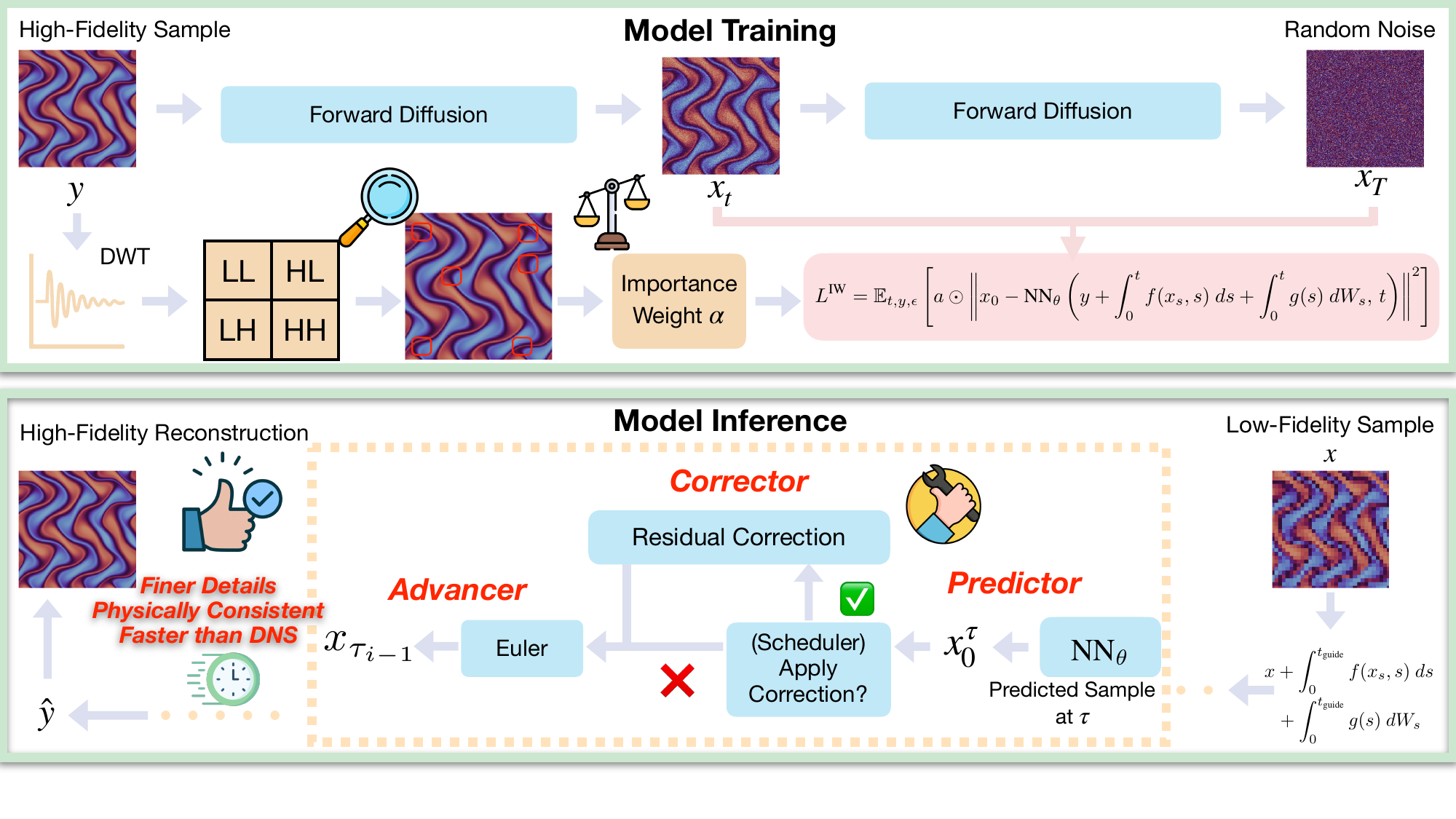}
    \caption{Training and inference pipeline of \model. Training with high-fidelity data only, guided by Importance Weight strategy to locate fine-grained high-fidelity details. During inference, low-fidelity data is used as guidance, and the Predictor-Corrector-Advancer solver ensures physical coherence.}
    \label{fig:framwork}
\end{figure*}

\section{Method:\model}
We build upon the diffusion framework proposed by \citet{shu2023physics} and study reconstructing high-fidelity data from solver-generated low-fidelity inputs, which have more information loss compared to artificially downsampled low-fidelity inputs. Our model, \model, features an \textit{Importance Weight} strategy that scores different components in the flow fields through the loss function in a self-supervised manner, forcing the model to recover more fine-grained high-fidelity details during training. In addition, a \textit{Predictor-Corrector-Advancer} SDE solver applies physics-informed correction during sampling, ensuring physical coherence in reconstructed samples during inference. The two modules jointly guide the model toward high-quality reconstruction from a wide range of low-fidelity inputs. The overall framework is depicted in Figure~\ref{fig:framwork}. We now introduce each component in detail.

\textbf{Model Setup.} \model~follows the guided data synthesis setting as in \citet{shu2023physics, meng2022sdedit}: we train the model with high-fidelity sources only, and condition on low-fidelity inputs as intermediate diffusion step during inference. This brings two benefits: (1) exerting control over the reconstruction process during inference. Instead of starting from random noises, the reverse diffusion starts from low-fidelity inputs as intermediate diffusion steps. (2) The model can reconstruct from generalized low-fidelity data, since the training does not depend on low- and high-fidelity pairs as in direct mapping models such as CNNs.

Formally, during the forward training process, we obtain intermediate diffusion states $x_t$ by solving the following forward SDE. We adopt VP-SDE \citep{ho2020denoising}, where $f(x,t) = -\frac{1}{2} \beta_t x_t$, $g(t) = \sqrt{\beta_t}$, and $\{ \beta_1, \ldots, \beta_T \}$ are variance schedules. The forward procedure continuously adds noises to the high-fidelity sample $x_0$.
\begin{align}
    dx_t = f(x_t,t) dt + g(t) dW_t,\ x_0 = y \sim p_{\mathcal{Y}}^{\text{train}}.
\end{align}

During inference stage, instead of starting from $x_T \sim N(0, I)$, we obtain the intermediate diffusion state $x_{t_{\text{guide}}} (0<t_{\text{guide}}<T)$ by adding noises to the conditioned low-fidelity data $x\sim p_{\mathcal{X}}^{\text{test}}$ as 
\begin{align} \label{eqn:transform}
    x_{t_{\text{guide}}} = x + \int_0^{t_{\text{guide}}} f(x_s, s)\ ds + \int_0^{t_{\text{guide}}} g(s)\ dW_s.
\end{align}
We adopt the same $t_{\text{guide}}$ as \citet{shu2023physics}, and use $x_{t_{\text{guide}}}$ as the starting point for the reverse process to reconstruct the high-fidelity sources. Then, we solve the reverse SDE in Equation~\ref{eqn:reverse_sde} to progressively produce a refined high-fidelity reconstruction that aligns with the low-fidelity conditioning data. \citet{shu2023physics} applied the above procedure to reconstruct high-fidelity data from artificially downsampled low-fidelity input. We demonstrate in Proposition~\ref{prop:dist} why their proposed diffusion model performs poorly for solver-generated low-fidelity data. For solver-generated low-fidelity data, the differences in noised probability distributions become larger, which hinders diffusion model's ability for accurate reconstruction. 

\begin{proposition}
\label{prop:dist}
Let $p_{\mathcal{Y}}$ denote the probability distribution of high-fidelity data. Let $p_{\mathcal{X}}$ and $q_{\mathcal{X}}$ denote the probability distribution of solver generated and downsampled low-fidelity data respectively. Define $\tilde{p}_{\mathcal{Y}}$, $\tilde{p}_{\mathcal{X}}$, and $\tilde{q}_{\mathcal{X}}$ to be the probability distribution transformed from $p_{\mathcal{Y}}$, $p_{\mathcal{X}}$, and $q_{\mathcal{X}}$ by applying Equation~\ref{eqn:transform}. Then, it follows that $D_{\text{KL}}(\tilde{p}_{\mathcal{Y}} \| \tilde{p}_{\mathcal{X}}) \geq D_{\text{KL}}(\tilde{p}_{\mathcal{Y}} \| \tilde{q}_{\mathcal{X}})$.
\end{proposition}

\subsection{Importance Weight During Training}\label{sec:importance_weight}
We therefore introduce an importance weighting mechanism during training as self-guidance to ensure accurate reconstruction of fine-grained details. Specifically, we transform the high-fidelity data into the wavelet domain using the DWT and assign importance scores to different components of fluid fields within the diffusion loss function. By considering spatial relationships between nearby vorticity, DWT has great abilities to locate information both in spatial and frequency domains~\citep{daubechies1992ten, akansu1992multiresolution}, thus allowing the model to capture fine-grained structures effectively. Formally, we decompose fluid fields $y\in\mathcal{R}^{n\times n}$ into frequency subdomains and compute the sum of squares of the high-frequency modes -- namely HL (high-low), LH (low-high) and HH (high-high) subdomains as below, where $HL, LH, HH, F \in \mathbb{R}^{\frac{n}{2} \times \frac{n}{2}}$. HL captures horizontal high-frequency signals, while LH captures vertical high-frequency signals. HH captures high-frequency signals in both directions, corresponding to diagonal details such as intersections.  
\begin{equation}
    F = HL^2 + LH^2 + HH^2.
\end{equation}
To compute importance scores, we then uniformly upsample $F$ to $\hat{F} \in \mathbb{R}^{n \times n}$ and linearly map $\hat{F}_{i,j}$ to an importance weight $a_{i,j}$ as below, where  $\alpha, \beta$ are the minimum and maximum importance weight value respectively, $Q_{\theta}(\hat{F}) \in \mathbb{R}$ is the $\theta$ quantile of all $\hat{F}$ values.
\begin{align}
    a_{i,j}
    =
    \begin{cases} 
        \alpha + (\beta - \alpha) \frac{\hat{F}_{i,j} - Q_{\theta}(\hat{F})}{\max \hat{F} - Q_{\theta}(\hat{F})} & \text{if } \hat{F}_{i,j} > Q_{\theta}(\hat{F}) \\
        1 & \text{otherwise}
\end{cases}
\end{align}
If $\hat{F}_{i,j}$ exceeds the $\theta$ quantile of all $\hat{F}$ values, the corresponding component is considered high-frequency, and will be assigned with weight greater than 1. Finally, the diffusion loss function in Eqn~\ref{eqn:loss_DDPM_original} is updated to incorporate the importance weighting. Note that our diffusion models directly predict the denoised sample instead of noise or score.
\begin{align}
    \resizebox{.92\linewidth}{!}{$
    L^{\text{IW}} = \mathbb{E}_{t, y, \epsilon} \left[ a \odot \left\| x_0 - \text{NN}_\theta \left( y + \int_0^{t} f(x_s, s)\, ds + \int_0^{t} g(s)\, dW_s,\, t \right) \right\|^2 \right],
    $}
\end{align}
where $x_0$ represents the ground truth high-fidelity sample,  $\text{NN}_{\theta}$ predicts the denoised high-fidelity sample, and $\odot$ denotes element-wise multiplication.

\textbf{Importance Weight Design Choice.}
Our \textit{importance weight} strategy is incorporated exclusively during training. The DWT-based calculation avoids the large computational complexity often associated with attention mechanisms. It efficiently emphasizes important features by leveraging the intrinsic properties of the wavelet transform, resulting in a more targeted learning process.

\textbf{Generalization of Hyperparameters}
The quantile $\theta$, and the minimum and maximum importance weight values $\alpha, \beta$ are tuned using only one dataset and generalize to all others.

\subsection{Predictor-Corrector-Advancer SDE Solver}\label{sec:residual}
In addition, we introduce a \textit{Predictor-Corrector-Advancer} SDE solver to enhance the physical coherence of the reconstructed data during inference. In the reverse diffusion process, the \textit{Predictor} first generate predicted high-fidelity data $\tilde{x}_0^{t} = \text{NN}_{\theta} (x_{t}, t)$. Then, the \textit{Corrector} applies residual corrections to $\tilde{x}_0^{t}$ to enhance physical consistency. Let $\mathcal{L}$ be a differential operator associated with a PDE acting on a function $u:\Omega \rightarrow \mathbb{R}^{n \times n}$, defined on a domain $\Omega \subset \mathbb{R}^{n \times n}$, with source term $f:\Omega \rightarrow \mathbb{R}^{n \times n}$. The PDE is expressed as $\mathcal{L}u(x) = f(x), \quad x \in \Omega$. Given an approximate solution $\tilde{u}(x)$, the PDE residual $\mathcal{R}(x)$ is defined as $\mathcal{R}(x) := || \mathcal{L} \tilde{u}(x) - f(x) ||_2^2$. The correction is to perform $M$ steps gradient descent based on the residuals using the Adam algorithm ~\citep{kingma2015adam}. Finally, the \textit{Advancer} leverages Euler–Maruyama \citep{maruyama1955continuous} to advance to the next diffusion step. The inference procedure is summarized in  Algorithm~\ref{algo:sample}.

\begin{algorithm*}
\caption{Predictor-Corrector-Advancer Inference Algorithm.}
\label{algo:sample}
\textbf{Require:} $x \in \mathcal{X}^{\text{test}}$, $t_{\text{guide}}$, $\tau = \{\tau_0, \tau_1, ..., \tau_K\}$ (backward steps, where $\tau_0 = 0, \cdots \tau_K = t_{\text{guide}}$), $\text{NN}_{\theta}$, $\mathcal{R}(\cdot)$, $M$ (gradient descent steps), $\eta$ (step size).
\begin{algorithmic}[1]
    \STATE $x_{\tau_k} = x + \int_0^{\tau_k} f(x_s, s)\ ds + \int_0^{\tau_k} g(s)\ dW_s.$
    \FOR{$i = K, K - 1, \dots, 1, 0$} 
        \STATE \textit{Predictor: } $\tilde{x}_0^{\tau_i} = \text{NN}_{\theta} (x_{\tau_i}, \tau_i)$
        \IF{correction is performed at time $\tau_{i-1}$}
            \REPEAT
                \STATE \textit{Corrector: } $\tilde{x}_0^{\tau_{i}} = \tilde{x}_0^{\tau_{i}} - \eta \cdot \text{Adam}(\nabla \mathcal{R}(\tilde{x}_0^{\tau_{i}}))$ 
            \UNTIL{$M$ times}
        \ENDIF

        \STATE \textit{Advancer: } $x_{\tau_{i-1}} = x_{\tau_{i}} + \left[  f(x_{\tau_{i}}, \tau_{i}) + g^2(\tau_{i}) \frac{x_{\tau_{i}} - \tilde{x}_0^{\tau_{i}} - \int_{0}^{\tau_{i}} f(x_s, s) ds}{\int_{0}^{\tau_{i}} g^2(s) \ dW_s} \right] dt + g(t) d\overline{W}_{\tau_{i}}$
    \ENDFOR
\end{algorithmic}
\end{algorithm*}

We begin by analyzing the convergence properties of the \textit{Corrector}. Proposition~\ref{prop:corrector_convergency} suggests that increasing the number of gradient descent steps generally leads to smaller residual values. Thus, for optimal physical consistency, Proposition~\ref{prop:corrector_convergency} suggests to perform more gradient descent steps, i.e. larger $M$.

\begin{proposition} [Regret Bound of \textit{Corrector} (\citet{kingma2015adam} Corollary 4.2)]
Define the regret $r(M) = \sum_{i=1}^M [\mathcal{R}(x_i) - \mathcal{R}(x^*)]$, where $x^* = \text{argmin}_{x} \sum_{i=1}^M \mathcal{R}(x)$. Assume $\mathcal{R}(x)$ has bounded gradient and distance between any $x_i$ generated by Adam is bounded. The \textit{Corrector} achieves the following guarantee, for all $M \geq 1$: $\frac{r(M)}{M} = O \left( \frac{1}{\sqrt{M}}\right)$.
\label{prop:corrector_convergency}
\end{proposition}

We then introduce the error bound for our \textit{Predictor-Corrector-Advancer} SDE solver in Proposition~\ref{prop:error_bound}, with proof detailed in Appendix~\ref{appendix:proof_error_bound}. The derived bound indicates that, under finite temporal discretization $\Delta t$, (1) an increase in the correction region, $|A|$, could lead to an increase in the prediction error; and (2) a higher number of gradient descent steps, $M$, similarly could result in larger predictive error. This observation is noteworthy because, while Proposition~\ref{prop:corrector_convergency} demonstrates that a larger $M$ enhances the physical consistency, Proposition~\ref{prop:error_bound} simultaneously reveals that it adversely affects predictive accuracy. Consequently, these findings highlight the importance of balancing the competing influences of physical consistency and prediction accuracy. To address such trade-off, Section~\ref{exp:scheduling} investigates into (1) determining an optimal schedule for applying the correction; and (2) establishing the optimal frequency of correction.


\begin{proposition}[Error Bound for \textit{Predictor-Corrector-Advancer}]
Let $\hat{x}_t$ denotes the piecewise solution of $x_t$ solved using the \textit{Predictor-Corrector-Advancer} method. Then,
\begin{align}
    Z(T) & = \sup_{0 \leq s \leq T} \mathbb{E} \left[ |x_s - \hat{x}_s |^2 \right]
    \leq O ( \Delta t \exp( O( |A|^2 (1 + \eta L_Q L_{\mathcal{R}})^{2M} )),
\end{align}
where $\eta, L_Q, L_{\mathcal{R}}$ are positive constant. $A$ denotes the region where correction is applied and $B$ otherwise. $A \cup B = [0, T]$ and $A \cap B = \emptyset$.
\label{prop:error_bound}
\end{proposition}

Additionally, we notice that the expected error approaches zero as the time discretization $\Delta t$ approaches zero. This implies the strong and weak convergency of our method.
\begin{proposition}[Strong and Weak Convergency of \textit{Predictor-Corrector-Advancer} SDE Solver]
The SDE solver outlined in Algorithm~\ref{algo:sample} satisfies both strong and weak convergency. Namely, for any appropriate test function $\phi$.
\begin{align}
    \lim_{\Delta t \rightarrow 0} \sup_{0 \leq s \leq T} \mathbb{E} \left[ |x_s - \hat{x}_s |^2 \right] & = 0 \quad \text{(Strong Convergency)} \\
    \lim_{\Delta t \rightarrow 0} \mid \mathbb{E} \left[ \phi(\hat{x}_{t_N})\right] - \mathbb{E} \left[ \phi(x_T)\right] \mid & = 0 \quad \text{(Weak Convergency)}
\end{align}
\label{prop:strong_weak_convergency}
\end{proposition}
The strong convergency implies the sample paths of the numerical approximation approach those of the true process in a mean sense. Equivalently, the error measured pathwise becomes arbitrarily small. Weak convergency is a result of strong convergency, and it guarantees the target distributions of the approximations converge.

\begin{table*}[!t] 
\centering
\scriptsize{
\resizebox{\linewidth}{!}{
    \setlength\tabcolsep{3pt}
    \renewcommand\arraystretch{1.4}
\begin{tabular}{l | l || ccccccc || ccc} 
\hline\hline 
\rowcolor{CadetBlue!20} 
& & DiffPIR & DPS & ResShift & PiRD & RSGD & Diff & Cond Diff & \cellcolor[HTML]{FFFFF0}\textbf{\model} & \cellcolor[HTML]{FFFFF0}\model\ w/o Cor & \cellcolor[HTML]{FFFFF0}\model\ w/o IW \\
\hline 

\multirow{4}{*}{\rotatebox{90}{\parbox{1.5cm}{\centering \textit{Taylor Green\\ Vortex}}}}
&\cellcolor{gray!15} L2   & \cellcolor{gray!15} 7.09 & \cellcolor{gray!15} 7.08 & \cellcolor{gray!15} 3.57 & \cellcolor{gray!15} 3.97 & \cellcolor{gray!15} 4.50 & \cellcolor{gray!15} 3.28 & \cellcolor{gray!15} 4.50 & \cellcolor[HTML]{FFFFF0}\textbf{3.18} & \cellcolor[HTML]{FFFFF0} 3.20 &\cellcolor[HTML]{FFFFF0} 3.22 \\
&\cellcolor{gray!15} Res. & \cellcolor{gray!15} 1434447.8194 & \cellcolor{gray!15} 1434333.5481 & \cellcolor{gray!15} 0.0426 & \cellcolor{gray!15} 0.9973 & \cellcolor{gray!15} 0.3892 & \cellcolor{gray!15} 51.5538 &\cellcolor{gray!15} 0.5392 & \cellcolor[HTML]{FFFFF0}\textbf{0.0016} &\cellcolor[HTML]{FFFFF0} 184.5116 & \cellcolor[HTML]{FFFFF0} 0.1306 \\
& L2                & 6.84 & 6.83 & 1.73 & 2.20 & 2.53 & 1.68  & 3.57 & \cellcolor[HTML]{FFFFF0}\textbf{1.55} & \cellcolor[HTML]{FFFFF0} 1.57 & \cellcolor[HTML]{FFFFF0} 1.59 \\
& Res.              & 1451893.9373 & 1451604.5279 & 0.1431 & 0.9912 & 7356.7753 & 25.0334 & 778.6521 & \cellcolor[HTML]{FFFFF0}\textbf{0.0196} & \cellcolor[HTML]{FFFFF0} 145.8101 & \cellcolor[HTML]{FFFFF0} 0.0377 \\
\hline 

\multirow{4}{*}{\rotatebox{90}{\parbox{1.5cm}{\centering \textit{Decaying Turbulence}}}}%
&\cellcolor{gray!15} L2   & \cellcolor{gray!15} 3.89  & \cellcolor{gray!15} 3.89 &\cellcolor{gray!15} 1.99  &\cellcolor{gray!15} 2.00 &\cellcolor{gray!15} 1.91  & \cellcolor{gray!15} 1.84 & \cellcolor{gray!15} 2.02 & \cellcolor[HTML]{FFFFF0}\textbf{1.71} & \cellcolor[HTML]{FFFFF0} 1.71 & \cellcolor[HTML]{FFFFF0} 1.72 \\
&\cellcolor{gray!15} Res. & \cellcolor{gray!15} 243726.9497 & \cellcolor{gray!15} 243724.2074 & \cellcolor{gray!15} 0.3971 & \cellcolor{gray!15} 0.9222 & \cellcolor{gray!15} 3178.7928 & \cellcolor{gray!15} 2111.2344 &\cellcolor{gray!15} 26592.0857 & \cellcolor[HTML]{FFFFF0}\textbf{0.0488} & \cellcolor[HTML]{FFFFF0} 13.2799 & \cellcolor[HTML]{FFFFF0} 0.1164 \\
& L2                & 3.51 & 3.51 & 0.93 & 0.96 & 0.87 & 0.85 & 1.34 & \cellcolor[HTML]{FFFFF0}\textbf{0.79} & \cellcolor[HTML]{FFFFF0} 0.81 & \cellcolor[HTML]{FFFFF0} 0.81 \\
& Res.              & 229694.9224 & 251006.8961 & 0.3935 & 0.9220 & 1363.2269 & 375.9871 & 4994.4697 & \cellcolor[HTML]{FFFFF0}\textbf{0.0059} & \cellcolor[HTML]{FFFFF0} 21.3210 & \cellcolor[HTML]{FFFFF0} 0.0143 \\
\hline 

\multirow{4}{*}{\rotatebox{90}{\parbox{1.5cm}{\centering \textit{Kolmogorov\\ Flow}}}}%
&\cellcolor{gray!15} L2   & \cellcolor{gray!15} 7.09 & \cellcolor{gray!15} 7.08 & \cellcolor{gray!15} 2.90 & \cellcolor{gray!15} 2.95 & \cellcolor{gray!15} 3.06 & \cellcolor{gray!15} 3.09 & \cellcolor{gray!15} 3.13 & \cellcolor[HTML]{FFFFF0}\textbf{2.78} & \cellcolor[HTML]{FFFFF0} 2.82 & \cellcolor[HTML]{FFFFF0} 2.93 \\
&\cellcolor{gray!15} Res. & \cellcolor{gray!15} 452127.1843 & \cellcolor{gray!15} 449397.5745 & \cellcolor{gray!15} 104.3518 & \cellcolor{gray!15} 0.8137 & \cellcolor{gray!15} 0.08677 & \cellcolor{gray!15} 9.5772 &\cellcolor{gray!15} 0.6837 & \cellcolor[HTML]{FFFFF0}\textbf{0.0082} & \cellcolor[HTML]{FFFFF0} 199.9087 & \cellcolor[HTML]{FFFFF0} 0.0803 \\
& L2                & 7.08 & 7.07 & 1.97 & 1.94 & 1.75 & 1.79 & 1.79 & \cellcolor[HTML]{FFFFF0}\textbf{1.61} & \cellcolor[HTML]{FFFFF0} 1.69 & \cellcolor[HTML]{FFFFF0} 1.73 \\
& Res.              & 452127.4891 & 449315.7908 & 95.9227 & 0.8754 & 1.3277 & 36.2915 & 7.6139 & \cellcolor[HTML]{FFFFF0}\textbf{0.0135} & \cellcolor[HTML]{FFFFF0} 41.1452 & \cellcolor[HTML]{FFFFF0} 0.3039 \\
\hline 

\multirow{4}{*}{\rotatebox{90}{\parbox{1.5cm}{\centering \textit{McWilliams\\ Flow}}}}%
&\cellcolor{gray!15} L2   & \cellcolor{gray!15} 4.37 & \cellcolor{gray!15} 4.36 & \cellcolor{gray!15} 2.27 & \cellcolor{gray!15} 2.28 & \cellcolor{gray!15} 2.33 & \cellcolor{gray!15} 2.23 & \cellcolor{gray!15} 2.24 & \cellcolor[HTML]{FFFFF0}\textbf{2.04} & \cellcolor[HTML]{FFFFF0} 2.06 &\cellcolor[HTML]{FFFFF0} 2.16 \\
&\cellcolor{gray!15} Res. & \cellcolor{gray!15} 54584.1591 & \cellcolor{gray!15} 54307.2485 & \cellcolor{gray!15} 0.1521 & \cellcolor{gray!15} 0.6638 & \cellcolor{gray!15} 0.5602 & \cellcolor{gray!15} 2.0254 & \cellcolor{gray!15} 0.8247 & \cellcolor[HTML]{FFFFF0}\textbf{0.0291} &\cellcolor[HTML]{FFFFF0} 79.9192 & \cellcolor[HTML]{FFFFF0} 0.0305 \\
& L2                & 4.36 & 4.36 & 1.44 & 1.43 & 1.45 & 1.29 & 1.30 & \cellcolor[HTML]{FFFFF0}\textbf{1.24} & \cellcolor[HTML]{FFFFF0} 1.27 & \cellcolor[HTML]{FFFFF0} 1.30 \\
& Res.              & 58511.5298 & 54284.0874 & 0.1068 & 0.6613 & 0.6371 & 5.0219 & 13.7145 & \cellcolor[HTML]{FFFFF0}\textbf{0.0017} & \cellcolor[HTML]{FFFFF0} 150.1363 & \cellcolor[HTML]{FFFFF0} 0.6516 \\
\hline 
\end{tabular}}}
\caption{Quantitative performance comparison over four datasets on L2 and Res metrics. Rows with a \colorbox{gray!15}{gray background} report results for 32$\times$32 $\to$ 256$\times$256 tasks, while rows with a white background report results for 64$\times$64 $\to$ 256$\times$256 tasks. Columns highlighted in \colorbox[HTML]{FFFFE0}{yellow} show results for our model and its ablations. \textbf{Bold} values indicate the best performance in each row.} 
\label{sr results}
\end{table*}

\textbf{Generalization of Correction Schedule}
The correction schedule and number of gradient descent steps are tuned using only one dataset and generalize to all others.

\textbf{Availability of PDE Residual} Our approach is designed to work in tandem with numerical PDE solvers, which require a complete specification of the PDE: its functional form, parameters, and boundary conditions. Consequently, it is reasonable to assume that the PDE residual is available.

\textbf{Contribution of Predictor-Corrector-Advance Solver}
While the use of PDE residuals in neural networks is well established, we introduce a novel mechanism for injecting them into diffusion models. \citet{huang2024diffusionpdegenerativepdesolvingpartial} and \citet{bastek2025physicsinformed} also embed PDE residuals in diffusion models. The former injects the residual’s gradient into the score, while the latter adds the residual to the loss function. \textcolor{red}{We instead apply several Adam updates to the predicted clean sample at selected reverse-diffusion steps. This design is motivated by the trade‑off between residual minimization and denoising performance as demonstrated by the error bound. Furthermore, in the experiment sections, we identify the scheduling policy that optimally balances these two objectives and demonstrate that it outperforms all baselines.} For comparison, we combine the residual score guidance of \citet{huang2024diffusionpdegenerativepdesolvingpartial} with the vanilla model of \citet{shu2023physics} to create a Residual Score-Guided Diffusion (RSGD) baseline and also include the residual-loss approach of \citet{shan2024pirdphysicsinformedresidualdiffusion} during experiments.

\subsection{Connection to Diffusion Posterior Sampling}
Diffusion models have shown great performance in image super-resolution tasks \citep{zhu2023denoising, chung2023diffusion, rozet2023scorebased, kawar2022denoising}. These studies assume that for high-fidelity data $y$ and low-fidelity data $x$, there exists $\mathcal{A}$ such that $x = \mathcal{A}(y) + \epsilon$, where $\epsilon$ is random noise.  While this holds for straightforward downsampling, our solver-generated low-fidelity data can only approximate by
\begin{align*}
     \resizebox{.95\linewidth}{!}{$
    \mathcal{A} (y) \approx \text{ForwardSolver}(\text{Downsample}(\text{ReverseSolver} (y, t_{\text{fluid}})), t_{\text{fluid}}).
    $}
\end{align*}
$\text{ReverseSolver} (y, t_{\text{fluid}}))$ integrates backward to estimate the initial high-fidelity state, which is downsampled and used by $\text{ForwardSolver}$ to produce the low-fidelity data. Computing gradients with respect to $A$ is impractical due to adaptive time stepping resulting in tens of thousands iterations for numerical solvers. As a result, previous approaches cannot be directly applied, but we still include some as baselines due to their prominence in related tasks. In our experiments, we adopt a downsampling strategy for $\mathcal{A}$ following \cite{chung2023diffusion}.

To incorporate low-fidelity data into the reverse diffusion process, we follow \citet{shu2023physics} by treating the low-fidelity data as intermediate reverse diffusion samples. We contend that this strategy offers stronger guidance. Although the gradient guidance from previous works facilitates projection onto the measurement subspace, our problem is characterized by a focus on a hard L2 loss, thereby requiring convergency to exact target point. As demonstrated in the experimental section, \citet{zhu2023denoising, chung2023diffusion} often produce significantly larger L2 loss.

\section{Experiments}
\subsection{Dataset}
We generate four 2D turbulent flow datasets using the incompressible Navier-Stokes: 1.) \textit{Taylor Green Vortex}, featuring gradually break down of large-scale vortices into smaller turbulent structures; 2.) \textit{Decaying Turbulence}, describing turbulence that evolves naturally without external forces; 3.) \textit{Kolmogorov Flow}, portraying turbulence influenced by a sinusoidal external force; 4.) \textit{McWilliams Flow}~\citep{Mcwilliams_1984}, describing the behavior of isolated vortices in turbulent conditions.
\begin{align*}
    \frac{\partial \omega(\bm{x}, t)}{\partial t} + \bm{u} (\bm{x}, t) \cdot \nabla \omega(\bm{x}, t) = \frac{1}{Re} \nabla^2 \omega(\bm{x}, t) + f(\bm{x}), \\
    \nabla \cdot \bm{u} (\bm{x}, t) = 0, \quad \omega(\bm{x}, 0) = \omega_0 (\bm{x}),
\end{align*}
where $\omega$ represents vorticity, $\bm{u}$ denotes velocity field, $Re$ is the Reynolds number, and $f(\bm{x})$ is an external forcing. $\omega_0$ represents the initial vorticity distribution. The PDE is numerically solved by pseudo-spectral solver~\citep{orszag1972comparison} on equispaced discretization grids. The high-fidelity data are generated with $4096 \times 4096$ grid and then uniformly downsampled to $256 \times 256$, while those on the lower-resolution grids are considered low-fidelity. The \textit{Taylor–Green Vortex} dataset contains 100 trajectories of 192 time steps at a Reynolds number of 1000. The \textit{Decaying Turbulence} dataset contains 400 trajectories of 64 time steps at a Reynolds number of 450. The \textit{Kolmogorov Flow} dataset contains 50 trajectories of 320 time steps at a Reynolds number of 1000. The \textit{McWilliams Flow} dataset contains 50 trajectories of 320 time steps at a Reynolds number of 2000. Each dataset is split with 80 percent of the trajectories for training, 10 percent for validation, and 10 percent for testing. We refer the readers to Appendix~\ref{appendix:dataset} for additional details.

\subsection{Experiment Settings}
\textbf{Task Setup and Baselines.} We evaluate two reconstruction settings:  $64 \times 64 \rightarrow 256 \times 256$ ($4 \times$ upsampling) and $32 \times 32 \rightarrow 256 \times 256$ ($8 \times$ upsampling). We compare against seven diffusion based reconstruction models: ResShift \citep{yue2023resshiftefficientdiffusionmodel}, PiRD \citep{shan2024pirdphysicsinformedresidualdiffusion}, DiffPIR \citep{zhu2023denoising}, Diffusion Posterior Sampling (DPS) \citep{chung2023diffusion}, Residual Score Guided Diffusion (RSGD) \citep{huang2024diffusionpdegenerativepdesolvingpartial}, Vanilla Diffusion (Diff) and its conditional variant (Cond Diff) from \citet{shu2023physics}. We perform two ablation studies, namely \model w/o IW and \model w/o Cor, where we remove \textit{Importance Weight} and \textit{Predictor-Corrector-Advancer} respectively. The comparisons between direct mapping models are available in \citet{shu2023physics} and Appendix~\ref{appendix:compare_direct_mapping}.

\textbf{Evaluation Metrics.} We assess the reconstructed flow fields using L2 norm for measuring the pointwise error and normalized residual errors (Res.) for assessing adherence to the underlying physics. According to \citet{shu2023physics}, lower residual values indicate enhanced physical consistency. Given that the high-fidelity residuals are also nonzero, we posit that reconstructed samples whose residuals more closely match those of the high-fidelity data are superior. The nonzero residuals observed in high-fidelity data can be attributed to the inherent approximations in residual computation with downsampled high-fidelity data, as well as the irreducible error present in numerical approximation. Formally, the metric is defined as $\frac{(\mathcal{R}(\hat{x}_0) - \mathcal{R}(y))^2}{\mathcal{R}(y)^2}$, where $\hat{x}_0$ is the predicted reconstruction and $y$ is the ground truth.

In addition, we conduct a novel multi-scale evaluation using DWT: we transform the predicted and ground truth flow fields into wavelet space and decompose them into four subdomains: LL, LH, HL, and HH. The LL subdomain captures large-scale, low-frequency information, while LH, HL, and HH encompass higher-frequency details like turbulent structures. By calculating the L2 norm in each subdomain, we gain a comprehensive understanding of the model's performance across different scales, ensuring accurate reconstruction of both global flow features and fine-scale details.

\begin{table}[H]
\centering
\begin{tabular}{lcccc}
\toprule
\textbf{Method} & \textbf{HH} & \textbf{HL} & \textbf{LL} & \textbf{LH} \\
\midrule
DiffPIR                & 4.3914  & 9.0811 & 188.6012  & 3.6674 \\
DPS                    & 4.3978  & 9.0713 & 187.9728  & 3.6626   \\
ResShift               & 0.1164  & 1.6399 & 13.4812 & 1.9845 \\
PiRD                   & 0.1093  & 1.6150 & 12.9042   & 1.9455 \\
RSGD                   & 0.0336  & 0.2134 & 13.5701 & 0.4348  \\
Diffusion              & 0.0279  & 0.2437 & 14.0867 & 0.4011 \\
Cond Diff  & 0.0294  & 0.2476 & 14.1269 & 0.4137 \\
\model          & \textbf{0.0263} & \textbf{0.2103} & \textbf{12.0338} & \textbf{0.3216} \\
\bottomrule
\end{tabular}
\caption{Kolmogorov Flow $4 \times$ Upsampling}
\label{tab:kolmogorov_u64}
\end{table}

\subsection{Reconstruction Results}\label{exp:reconstruction}
Table~\ref{sr results} reports the L2 loss and PDE residual across datasets and models. For both upsampling scales, \model~consistently outperforms baselines, showing its effectiveness. The lower L2  achieved by \model~indicates that it effectively captures essential features and dynamics of the turbulent flows. Additionally, residuals from \model~reconstructions are closer to those of the high-fidelity data. The small difference in PDE residuals indicates that \model's predictions adhere more closely to the underlying physical laws. For complex datasets dominated by fine-grained details, such as \textit{Kolmogorov Flow} and \textit{McWilliams Flow}, \model\ outperforms baselines by a margin, showing better reconstruction accuracy and physical coherence. Through ablation studies, we demonstrate that both \textit{Importance Weight} and \textit{Predictor-Corrector-Advancer} contribute significantly to improving model's performance, underscoring the effectiveness of our design choices in capturing the complex behaviors of turbulent flows. DiffPIR and DPS have large L2 loss and PDE residual, as expected. This can be attributed to using downsampling to approximate $\mathcal{A}$ and the weaker guidance imposed by these two methods. Although PiRD yields very small PDE residuals for \textit{Taylor–Green Vortex} and \textit{Decaying Turbulence}, these residuals still differ substantially from the high-fidelity reference data. In other words, the method prioritizes reducing the PDE residual at the expense of predictive accuracy, and the smaller residual does not translate into better flow reconstruction.

\textbf{Explanation on Large Residual.} 
When you evaluate a PDE residual on a predicted solution, even tiny inaccuracies in your approximation get blown up by the nature of differentiation: taking derivatives accentuates any small noise in the predicted field, turning minute errors into large swings. Further, every PDE comes with an implicit requirement on how many times its solution must be differentiable, so when the predicted flow fields are only piecewise continuous,  any discontinuity or abrupt change in slope gets turned into a large error.

\textbf{Multi-Scale Evaluation.} 
We assess the model's ability to capture flow structures at different scales using DWT. We present the results on 4$\times$ upsampling case in \textit{Kolmogorov Flow} in Table~\ref{tab:kolmogorov_u64} and additional results in Appendix~\ref{appendix:multi_scale_eval}. \model\ demonstrates superior performance in all subdomains. Excelling in the LL subdomain indicates a strong capability in capturing large-scale, low-frequency components of the turbulent flows. The superior performance in the LH and HL subdomains suggests the effectiveness of \model~in capturing small-scale vortices and transitions between scales.

\textbf{Runtime Comparison.}
We report the runtime comparison of different methods in Appendix~\ref{appendix:runtime}. \model~only increases small inference time while achieving superior performance improvement against baselines. The total runtime of using the numerical solvers to generate low-fidelity data, and reconstructing with \model~is considerably faster than DNS. Note that numerical solvers employs adaptive time stepping governed by Courant–Friedrichs–Lewy (CFL) condition \citep{courant1928uber}.


\textbf{Sensitivity Analysis.}
We study the effects of three key hyperparameters in calculating the importance weight: the maximum importance weight $\beta$, minimum importance weight $\alpha$, and the threshold parameter $\theta$. The results in Appendix~\ref{appendix:ses_imw} show that increasing $\beta$ reduces both L2 and PDE residuals, indicating a broader range of importance weights is beneficial. Smaller $\alpha$ values improve performance while setting $\alpha$ to 1 is suboptimal. This can be understood as it decreases the difference between high-frequency and low-frequency regions, as the weight for the latter one is set to 1. Additionally, a larger $\theta$ (0.7 or 0.8) helps the model focus on important details. The experiments presented in Appendix~\ref{appendix:ses_imw} are conducted on the validation datasets for \textit{Kolmogorov Flow}. The optimal hyperparameters, selected based on this validation dataset, are applied across all datasets during testing.

\begin{table}[H]
\centering
\small
\begin{tabular}{lccc|ccc}
\toprule
\textbf{Model} & \multicolumn{3}{c|}{\textbf{4$\times$ Upsampling}} & \multicolumn{3}{c}{\textbf{8$\times$ Upsampling}} \\
\cmidrule(lr){2-4}\cmidrule(lr){5-7}
& \textbf{PSNR} & \textbf{SSIM} & \textbf{LPIPS} & \textbf{PSNR} & \textbf{SSIM} & \textbf{LPIPS} \\
\midrule
DiffPIR     & 22.6139 & 0.0512 & 0.2318 & 21.5010 & 0.0394 & 0.6607 \\
DPS         & 22.7173 & 0.0530 & 0.2148 & 22.7461 & 0.0532 & 0.6606 \\
ResShift    & 24.4684 & 0.5587 & 0.1350 & 19.1656 & 0.3099 & 0.3343 \\
PiRD        & 23.4082 & 0.5378 & 0.1625 & 19.8004 & 0.3242 & 0.3657 \\
RSGD        & 25.0200 & 0.6416 & 0.1233 & 20.6228 & 0.3608 & 0.2915 \\
Diffusion   & 25.4049 & 0.6487 & 0.1215 & 21.5818 & 0.4072 & 0.2848 \\
Cond Diff   & 25.2389 & 0.6456 & 0.1229 & 20.2067 & 0.3425 & 0.2869 \\
\model      & \textbf{26.1733} & \textbf{0.6781} & \textbf{0.1097} & \textbf{24.0754} & \textbf{0.4409} & \textbf{0.2781} \\
\bottomrule
\end{tabular}
\caption{PSNR, SSIM and LPIPS scores on \textit{Kolmogorov Flow} for 4× and 8× upsampling.}
\label{tab:metrics_klm}
\end{table}

\textbf{Image Similarity Metrics. } We also evaluate the mode on Learned Perceptual Image Patch Similarity (LPIPS) score \citep{zhang2018unreasonable}, Peak Signal-to-Noise Ratio (PSNR) \citep{Gonzalez2002}, and Structural Similarity Index Measure (SSIM) \citep{Wang2004}. We present the results on \textit{Kolmogorov Flow} in Table~\ref{tab:metrics_klm}. Additional results on other datasets are presented in Appendix~\ref{appendix:sim_metrics}. Results show that \model~demonstrate superior performance in these metrics under all settings.

\textbf{Model Generalization. } We observe that \model~generalizes well to low-fidelity data generated with different solver configurations. The experiment results are reported in Appendix~\ref{appendix:transfer}. We generate new \textit{Kolmogorov Flow} datasets by varying solver configurations. We compare our model trained with the original \textit{Kolmogorov Flow} datasets with our model trained directly on the new datasets. Results show that our model trained on the original datasets has comparable performance.

\subsection{Physical Guidance}
\label{exp:scheduling}
We conduct systematic study on how to schedule the corrector. The experiments are conducted on the validation datasets for \textit{Kolmogorov Flow}. The optimal hyperparameters, selected based on this validation dataset, are applied across all datasets during testing.
\begin{table}[H]
\centering
\small
\scalebox{1.0}{
\begin{tabular}{lcccc}
\toprule
\multirow{2}{*}{Schedule} & \multicolumn{2}{c}{32$\times$32 $\rightarrow$ 256$\times$256} & \multicolumn{2}{c}{64$\times$64 $\rightarrow$ 256$\times$256} \\
\cmidrule(lr){2-3} \cmidrule(lr){4-5}
& L2  & Res. & L2  & Res. \\
\midrule
Uniform 4         & 2.7910   & 0.1628    & 1.6645   & 0.0226  \\
\hline 
Start 3 End 1     & 2.7906   & 0.5913    & 1.6617   & 0.1206  \\
Start 2 End 2     & 2.7897   & 0.4391    & 1.6609   & 0.4309  \\
Start 1 End 3     & 2.7896   & 2.183    & 1.6617   & 2.1352  \\
\hline
Start 4 Space 1   & \textbf{2.7889}   & 10.3074   & \textbf{1.6587}   & 11.4666 \\
Start 4 Space 2   & \textbf{2.7888}   & 9.4531    & \textbf{1.6590}   & 10.4516 \\
Start 4 Space 3   & \textbf{2.7887}   & 9.4531    & \textbf{1.6594}   & 8.6455  \\
\hline
End 4 Space 1     & 2.7930   & \textbf{0.0400}    & 1.6668   & \textbf{0.0080}  \\
End 4 Space 2     & 2.7932   & \textbf{0.0743}    & 1.6747   & \textbf{0.0172}  \\
End 4 Space 3     & 2.7924   & \textbf{0.0407}    & 1.6730   & \textbf{0.0031}  \\
\bottomrule
\end{tabular}
}
\caption{Scheduling policy comparison on \textit{Kolmogorov Flow}. Applying corrector at the end leads to a reduced PDE residual, while placing it at the beginning achieves a lower L2 loss.}
\label{table:ResidualCorrectionPlacement}
\end{table}

\textbf{Scheduling Policy.}
We first explore the optimal schedule policy by comparing: 1.) \textit{Uniform N}: Distributing N residual correction steps evenly across diffusion steps; 2.) \textit{Start I, End N} Placing I consecutive correction steps at the start and N at the end of the diffusion process; 3.) \textit{Start N, Space S}: Placing N correction steps at the start with a spacing of S; 4.) \textit{End N, Space S}: Placing N correction steps at the end with a spacing of S. Results in Table~\ref{table:ResidualCorrectionPlacement} 
suggest that applying correction steps at the beginning achieves the lowest L2, while applying at the end has the lowest PDE residuals.
To balance between L2 and PDE residual, we adopt the \textit{Start N End N} schedule.

\begin{figure}[htbp]
    \centering
    \scalebox{0.95}{
    \begin{minipage}{0.95\linewidth}
        \centering
        \begin{subfigure}
            \centering
            \includegraphics[width=0.47\linewidth]{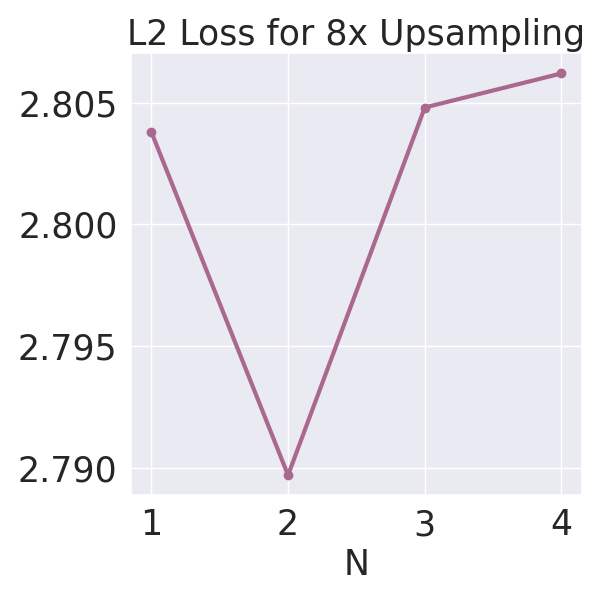}
        \end{subfigure}
        \hfill
        \begin{subfigure}
            \centering
            \includegraphics[width=0.47\linewidth]{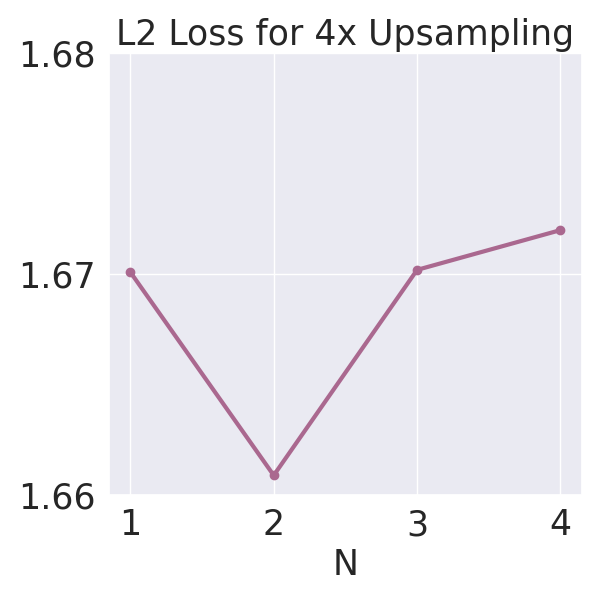}
        \end{subfigure}

        \vspace{1em}

        \begin{subfigure}
            \centering
            \includegraphics[width=0.47\linewidth]{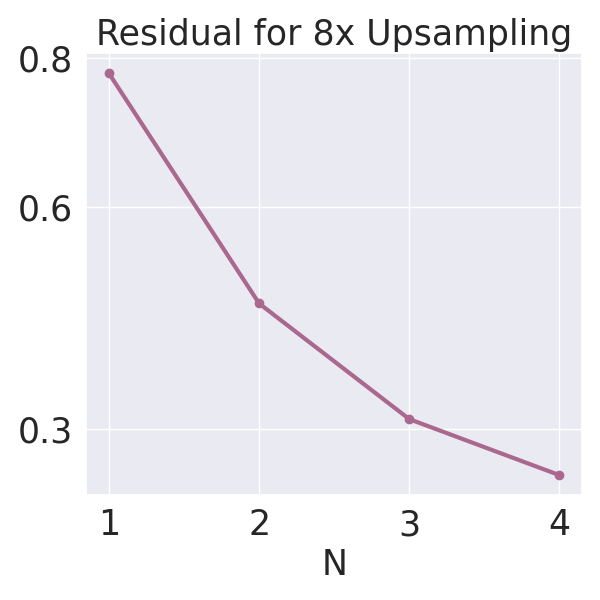}
        \end{subfigure}
        \hfill
        \begin{subfigure}
            \centering
            \includegraphics[width=0.47\linewidth]{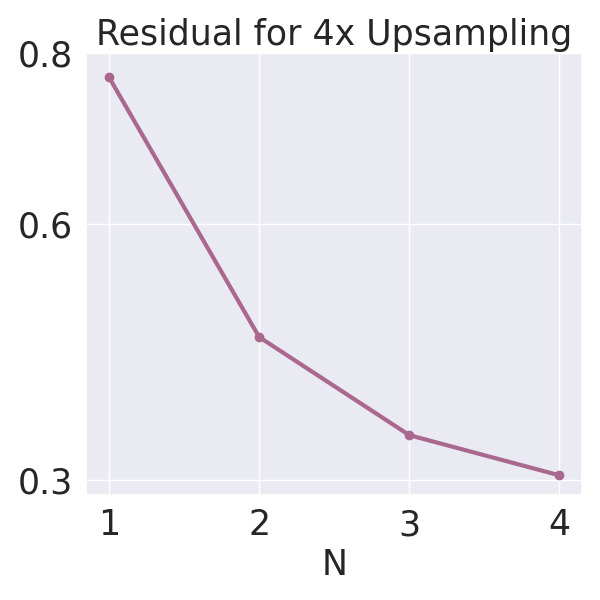}
        \end{subfigure}
    \end{minipage}
    }
    \caption{Varying number of $N$ using the \textit{Start N, End N} correction schedule on the \textit{Kolmogorov Flow} dataset. Showing both L2 and Residual across upsampling settings.}
    \label{fig:res_cor_steps}
\end{figure}

\textbf{Number of Correction Steps.}
We vary the number of $N$ in the \textit{Start N, End N} policy as shown in Figure~\ref{fig:res_cor_steps}. We observe that increasing $N$ leads to enhanced physical coherence measured by PDE residuals.
However, L2 does not continuously decrease with larger $N$. It reaches a minimum when $N=2$. This suggests that while more correction steps improve models' adherence to physical laws, an excessive number may interfere with the model's ability to accurately capture the intricate details of the turbulent flow. 
Therefore, we use \textit{Start 2, End 2} as the optimal balance.

\section{Conclusion}
We study a novel problem of reconstructing high-fidelity CFD from solver-generated low-fidelity inputs. Our diffusion model, \model, achieves high-quality reconstruction guided by \textit{Importance Weight} and \textit{Predictor-Corrector-Advancer} solver.

\section{Limitations}
Although the proposed diffusion framework achieves a significant speedup compared to high-fidelity solvers, it remains relatively slow compared to direct mapping models. Future work could explore recent advancements in accelerating diffusion sampling, such as \citet{lu2022dpm}, to further enhance the efficiency of the reconstruction process.

\section{Broader Impacts}
By enhancing the accuracy of CFD simulations, this approach can significantly reduce the computational costs associated with high-resolution simulations, which are often prohibitively expensive in terms of both time and resources. This has implications for a wide range of industries, including aerospace, automotive design, and environmental engineering, where high-fidelity simulations are essential for optimizing performance, safety, and sustainability. Additionally, by improving the quality of low-resolution CFD data, our work could enable more accessible, efficient research and development processes, allowing smaller organizations and research teams to leverage advanced simulation capabilities.

\section{Acknowledgments}
This work was partially supported by National Artificial Intelligence Research Resource (NAIRR) Pilot (240280, 240443), National Science Foundation (2106859, 2211557, 2119643, 2200274, 2303037, 2312501, 2531008), National Institutes of Health (U54HG012517, U24DK097771, U54OD036472, OT2OD038003, R01HL175135), NEC, Optum AI, SRC JUMP 2.0 Center, Amazon Research Awards, and Snapchat Gifts.

\bibliographystyle{ACM-Reference-Format}
\bibliography{sample-base}

\appendix



\clearpage
\newpage

\section{Dataset}
\label{appendix:dataset}
All datasets are generated using pseudo-spectral solver. The time-stepping method employed is a combination of the Crank-Nicholson scheme and Heun’s method. For numerical stability, the solver uses an adaptive time-stepping approach governed by the Courant-Friedrichs-Lewy (CFL) condition. The CFL condition ensures that the time step remains sufficiently small relative to the velocity field and the external forcing, preventing instabilities that can arise from rapid changes in the solution. Additionally, the dealiasing procedure, using the 2/3 rule, removes high-frequency components from the Fourier spectrum, ensuring that non-physical aliasing effects are avoided.

\begin{algorithm}
\caption{Pseudo-spectral Navier-Stokes Solver}
\begin{algorithmic}[1]
\STATE \textbf{Input:} Initial vorticity $\omega_0$, forcing $f$, Reynolds number $Re$, total time $T$, timestep $\Delta t$, domain size $L_1$, $L_2$, grid size $s_1$, $s_2$, adaptivity flag
\STATE \textbf{Output:} Vorticity $\omega$ at time $T$
\STATE \textbf{Initialize:} Compute wavenumbers, Laplacian, and dealiasing mask
\STATE \WHILE{$t < T$}
\IF{adaptive}
    \STATE Compute velocity field $\bm{u} = \nabla^\perp \psi$
    \STATE Update timestep $\Delta t$ based on CFL
\ENDIF
\STATE Compute non-linear term in Fourier space
\STATE Predictor and corrector steps using Crank-Nicholson + Heun
\STATE Apply dealiasing mask
\STATE Update time $t = t + \Delta t$
\ENDWHILE
\STATE Return Vorticity $\omega$
\end{algorithmic}
\end{algorithm}

\textbf{Taylor Green Vortex} The initial vorticity field is based on the analytical solution of the TGV, and to generate different trajectories, we added random perturbations from a Gaussian random field. These perturbations introduce variability to the initial conditions while maintaining the overall vortex structure. No external forcing was applied during the simulation, and the spatial domain is $[0, \frac{3}{2} \pi]^2$with a fixed Reynolds number of $1000$. The simulation used a time step $dt = \frac{1}{32}$, and $100$ trajectories were generated each with a total duration of $T = 6$ seconds. The initial vorticity field is based on the analytical solution for the two-dimensional periodic domain.
The vorticity field $\omega$ is initialized as $\omega = -2U_0 k \sin(kx) \sin(ky)$, where $U_0$ is the initial velocity amplitude, and $k$ is the wave number that determines the size of the vortices. To introduce variability and generate different trajectories, a Gaussian Random Field is added as a perturbation to the initial vorticity.

\textbf{Decaying Turbulence} The spatial domain is \( [0, 1]^2 \), with periodic boundary conditions and a fixed Reynolds number of 450. The simulation used a time step \( dt = \frac{1}{32} \), and 400 trajectories were generated, each with a total duration of \( T = 2 \) seconds. The initial conditions for the decaying turbulence dataset are generated by superimposing randomly positioned vortices of varying intensity and size. Each vortex is characterized by a randomly selected core size and maximum rotational velocity, allowing for a diverse range of initial flow structures. The vortices are distributed randomly throughout the domain, and their periodic images are added to ensure the proper enforcement of periodic boundary conditions.

\textbf{Kolmogorov Flow} The initial vorticity field is generated using a Gaussian random field, and the system is subjected to a forcing term of the form $f(\bm{x}) = -4 \cos(4x_2) - 0.1 \omega(\bm{x}, t)$. This forcing drives the flow in the y-direction while introducing a drag force that dissipates energy. The spatial domain is $ [0, 2\pi]^2 $, with a fixed Reynolds number of 1000. The simulation used a time step $dt = \frac{1}{32}$, and 50 trajectories were generated, each with a total duration of $T = 10$ seconds. The initial vorticity field is produced by sampling from a Gaussian Random Field. As the external forcing continually adds energy to the system, the initially simple vorticity evolves into intricate and turbulent structures. The vorticity is allowed to evolve over a 5-second period, and the final state at the end of this interval is used as the initial condition for our dataset.

\textbf{McWilliams Flow}
The phenomenon illustrates the emergence of order from initially disordered turbulent motion, driven by viscous dissipation and the self-organization of the flow. No external forcing is applied during the simulation, allowing for a natural decay and evolution of the turbulence. The spatial domain is \( [0, 2\pi]^2 \), with periodic boundary conditions and a Reynolds number of 2000, providing a high degree of turbulence. The simulation used a time step \( dt = \frac{1}{32} \), and 50 trajectories were generated, each with a total duration of \( T = 10 \) seconds. The initial vorticity field for the \textit{McWilliams Flow} is generated following the method described by \citet{Mcwilliams_1984}. The process begins by constructing a Fourier mesh over the spatial domain, where the wavenumbers $k_x$ and $k_y$ are calculated. A scalar wavenumber function is prescribed, and the ensemble variance is determined to ensure that the energy distribution in Fourier space follows the desired spectral shape. Random Gaussian perturbations are applied to each Fourier component of the stream function, producing a random realization of the vorticity field. To ensure the stream function has a zero mean, a spectral filter is applied, and the field is normalized based on the kinetic energy. Finally, the vorticity field is computed in physical space by taking the inverse Laplacian of the stream function in Fourier space, resulting in a turbulent flow field that evolves naturally without external forcing.


\section{Experiment Results for Direct Mapping Models}
\label{appendix:compare_direct_mapping}
For direct mapping models, we compare against a CNN-based model \citep{Fukami_Fukagata_Taira_2019}, a GAN-based model \citep{Generative}. Note that we apply 5 steps of gradient descent using the residual to the reconstructed sample to enhance physical consistency. The results are reported in Table~\ref{tab:direct_mapping_results}.

\begin{table}[H]
\centering
\small 
\setlength{\tabcolsep}{2pt}
\scalebox{0.8}{
\begin{tabular}{lccc}
\toprule
 & Metric & CNN & GAN \\
\midrule
\multirow[c]{5}{*}{\rotatebox{90}{\parbox{2cm}{\centering \textit{Taylor Green Vortex}}}}
 & \cellcolor{gray!25}{L2} & \cellcolor{gray!25} 3.66 & \cellcolor{gray!25} 3.53  \\
 & \cellcolor{gray!25}{Res.} & \cellcolor{gray!25} 139.7468 & \cellcolor{gray!25} 1.7611 \\
\\
 & L2              & 2.69 & 0.9990 \\
 & Res.            & 52661.27 & 0.7642 \\
 \midrule
 \multirow[c]{5}{*}{\rotatebox{90}{\parbox{2cm}{\centering \textit{Decaying Turbulence}}}}
 & \cellcolor{gray!25}{L2} & \cellcolor{gray!25} 2.01 & \cellcolor{gray!25} 2.09  \\
 & \cellcolor{gray!25}{Res.} & \cellcolor{gray!25} 82.1334 & \cellcolor{gray!25} 6.7948 \\
\\
 & L2              & 1.40 & 2.04 \\
 & Res.            & 57.1200 & 4.1499 \\
  \midrule
 \multirow[c]{5}{*}{\rotatebox{90}{\parbox{2cm}{\centering \textit{Kolmogorov Flow}}}}
 & \cellcolor{gray!25}{L2} & \cellcolor{gray!25} 3.41 & \cellcolor{gray!25} 3.11  \\
 & \cellcolor{gray!25}{Res.} & \cellcolor{gray!25} 515.8137 & \cellcolor{gray!25} 69.5502 \\
\\
 & L2              & 3.13 & 3.04 \\
 & Res.            & 185.4361 & 62.6224 \\
  \midrule
 \multirow[c]{5}{*}{\rotatebox{90}{\parbox{2cm}{\centering \textit{McWilliams Flow}}}}
 & \cellcolor{gray!25}{L2} & \cellcolor{gray!25} 2.31 & \cellcolor{gray!25} 2.20  \\
 & \cellcolor{gray!25}{Res.} & \cellcolor{gray!25} 654.7895 & \cellcolor{gray!25} 1533.1234 \\
\\
 & L2              & 1.61 & 1.92 \\
 & Res.            & 712.5844 & 1540.6807 \\
\bottomrule
\end{tabular}}
\caption{Quantitative performance comparison over four datasets on L2 and Res. Metrics are reported for both 32×32 → 256×256 (\colorbox{gray!25}{gray}) and 64×64 → 256×256 tasks.}
\label{tab:direct_mapping_results}
\end{table}

\section{Multi-Scale Evaluation}
\label{appendix:multi_scale_eval}

\begin{table}[H]
\centering
\caption{Kolmogorov Flow $8 \times$ Upsampling}
\label{tab:kolmogorov_u64}
\begin{tabular}{lcccc}
\toprule
\textbf{Method} & \textbf{HH} & \textbf{HL} & \textbf{LL} & \textbf{LH} \\
\midrule
DiffPIR                & 4.3914  & 9.0811 & 188.6012  & 3.6674 \\
DPS                    & 4.3914  & 9.0795 & 188.0532  & 3.6646   \\
ResShift               & 0.0722  & 1.8942 & 39.4973 & 2.1263 \\
PiRD                   & 0.0649  & 1.8410 & 37.2648 & 2.0639 \\
RSGD                   & 0.0346  & 0.2947 & 40.0763 & 0.4904  \\
Diffusion              & 0.0382  & 0.2935 & 41.8237 & 0.4429 \\
Cond Diff  & 0.0344  & 0.2903 & 42.9040 & 0.4445 \\
\model          & \textbf{0.0312} & \textbf{0.2616} & \textbf{33.8086} & \textbf{0.3798} \\
\bottomrule
\end{tabular}
\end{table}

\begin{table}[H]
\centering
\caption{McWilliams Flow $4 \times$ Upsampling}
\label{tab:kolmogorov_u64}
\begin{tabular}{lcccc}
\toprule
\textbf{Method} & \textbf{HH} & \textbf{HL} & \textbf{LL} & \textbf{LH} \\
\midrule
DiffPIR                & 0.4226  & 1.1473 & 75.6783  & 0.5940 \\
DPS                    & 0.4225  & 1.1465 & 75.4513  & 0.5932   \\
ResShift               & 0.0523  & 0.9234 & 6.8736 & 0.8362 \\
PiRD                   & 0.0516  & 0.9271 & 6.8646 & 0.8350 \\
RSGD                   & 0.0203  & 0.2155 & 6.6627 & 0.1959  \\
Diffusion              & 0.0182  & 0.2108 & 6.7780 & 0.1834 \\
Cond Diff  & 0.0167  & 0.2229 & 6.8123 & 0.1944 \\
\model          & \textbf{0.0141} & \textbf{0.1642} & \textbf{6.2937} & \textbf{0.1452} \\
\bottomrule
\end{tabular}
\end{table}

\begin{table}[H]
\centering
\caption{McWilliams Flow $8 \times$ Upsampling}
\label{tab:kolmogorov_u64}
\begin{tabular}{lcccc}
\toprule
\textbf{Method} & \textbf{HH} & \textbf{HL} & \textbf{LL} & \textbf{LH} \\
\midrule
DiffPIR                & 0.4182  & 1.1362 & 75.4749  & 0.6381 \\
DPS                    & 0.4225  & 1.1468 & 75.4864  & 0.5934   \\
ResShift               & 0.0320  & 0.9873 & 19.3344 & 0.9170 \\
PiRD                   & 0.0321  & 0.9917 & 19.4258 & 0.9253 \\
RSGD                   & 0.0152  & 0.2288 & 23.2287 & 0.1983  \\
Diffusion              & 0.0152  & 0.2287 & 20.9631 & 0.2027 \\
Cond Diff  & 0.0151  & 0.2353 & 21.1752 & 0.2085 \\
\model          & \textbf{0.0145} & \textbf{0.1920} & \textbf{17.4284} & \textbf{0.1708} \\
\bottomrule
\end{tabular}
\end{table}

\begin{table}[H]
\centering
\caption{Decaying Turbulence $4 \times$ Upsampling}
\label{tab:kolmogorov_u64}
\begin{tabular}{lcccc}
\toprule
\textbf{Method} & \textbf{HH} & \textbf{HL} & \textbf{LL} & \textbf{LH} \\
\midrule
DiffPIR                & 5.9452  & 9.0297 & 35.3487  & 3.7063 \\
DPS                    & 5.9452  & 9.0292 & 35.3019  & 3.7059   \\
ResShift               & 0.0504  & 0.7032 & 5.1194 & 0.7208 \\
PiRD                   & 0.0500  & 0.7140 & 5.3197 & 0.7326 \\
RSGD                   & 0.0442  & 0.0611 & 7.4907 & 0.0431  \\
Diffusion              & 0.0765  & 0.0995 & 5.1389 & 0.0994 \\
Cond Diff  & 0.0365  & 0.0510 & 29.6782 & 0.0546 \\
\model          & \textbf{0.0196} & \textbf{0.0264} & \textbf{5.0720} & \textbf{0.0267} \\
\bottomrule
\end{tabular}
\end{table}

\begin{table}[H]
\centering
\caption{Decaying Turbulence $8 \times$ Upsampling}
\label{tab:kolmogorov_u64}
\begin{tabular}{lcccc}
\toprule
\textbf{Method} & \textbf{HH} & \textbf{HL} & \textbf{LL} & \textbf{LH} \\
\midrule
DiffPIR                & 5.9592  & 9.0601 & 52.0273  & 3.7367 \\
DPS                    & 5.9592  & 9.0600 & 52.0212  & 3.7366   \\
ResShift               & 0.0416  & 1.1680 & 26.8325 & 1.2171 \\
PiRD                   & 0.0416  & 1.1843 & 26.7621 & 1.2353 \\
RSGD                   & 0.0425  & 0.0508 & 29.0300 & 0.0896  \\
Diffusion              & 0.1780  & 0.2204 & 25.0249 & 0.2159 \\
Cond Diff  & 0.0461  & 0.0511 & 34.5407 & 0.0591 \\
\model          & \textbf{0.0238} & \textbf{0.0488} & \textbf{24.1640} & \textbf{0.0512} \\
\bottomrule
\end{tabular}
\end{table}

\begin{table}[H]
\centering
\caption{Taylor Green Vortex $4 \times$ Upsampling}
\label{tab:kolmogorov_u64}
\begin{tabular}{lcccc}
\toprule
\textbf{Method} & \textbf{HH} & \textbf{HL} & \textbf{LL} & \textbf{LH} \\
\midrule
DiffPIR                & 19.4488  & 42.9432 & 127.4038  & 5.5207 \\
DPS                    & 19.4477  & 42.9413 & 127.1114  & 5.5192   \\
ResShift               & 0.3451  & 1.3825 & 13.0696 & 1.3727 \\
PiRD                   & 0.3422  & 1.3822 & 20.4812 & 1.3618 \\
RSGD                   & 0.3385  & 0.5539 & 27.0561 & 0.4622  \\
Diffusion              & 0.1703  & 0.2959 & 11.3933 & 0.2928 \\
Cond Diff  & 0.2941  & 0.5003 & 77.4125 & 0.4899 \\
\model          & \textbf{0.1507} & \textbf{0.2928} & \textbf{9.3202} & \textbf{0.2903} \\
\bottomrule
\end{tabular}
\end{table}

\begin{table}[H]
\centering
\caption{Taylor Green Vortex $8 \times$ Upsampling}
\label{tab:kolmogorov_u64}
\begin{tabular}{lcccc}
\toprule
\textbf{Method} & \textbf{HH} & \textbf{HL} & \textbf{LL} & \textbf{LH} \\
\midrule
DiffPIR                & 19.4647  & 42.9816 & 141.3724  & 5.5507 \\
DPS                    & 19.4671  & 42.9863 & 141.0834  & 5.5499   \\
ResShift               & 0.2900  & 1.9425 & 44.7676 & 1.9238 \\
PiRD                   & 0.2811  & 1.9054 & 66.0041 & 1.8890 \\
RSGD                   & 0.0284  & 0.4015 & 84.0803 & 0.3705  \\
Diffusion              & 0.0998  & 0.3941 & 44.8146 & 0.3727 \\
Cond Diff  & 0.0174  & 0.3832 & 81.7034 & 0.3772 \\
\model          & \textbf{0.0172} & \textbf{0.3811} & \textbf{41.5014} & \textbf{0.3704} \\
\bottomrule
\end{tabular}
\end{table}

\section{Image Similarity Metrics}
\label{appendix:sim_metrics}
Although LPIPS is trained on the ImageNet dataset, which consists of natural images, it remains a valuable metric for evaluating perceptual quality in CFD applications. This is because LPIPS leverages features from deep neural networks that are effective at capturing multi-scale patterns, textures, and perceptual similarities, regardless of the specific domain. Fluid dynamics data often have complex structures and turbulent patterns that share characteristics with textures found in natural images, making LPIPS suitable for assessing the fidelity of reconstructed flow fields. Thus, despite being trained on ImageNet, LPIPS can still effectively quantify how well the reconstructed samples retain important perceptual details, making it a robust metric for evaluating the visual quality of CFD reconstructions.

\begin{table}[H]
\centering
\small
\begin{tabular}{lcc}
\toprule
Model & 8$\times$ Upsampling & 4$\times$ Upsampling \\
\midrule
DiffPIR     & 0.5993 & 0.2139 \\
DPS         & 0.5993 & 0.2148 \\
ResShift    & 0.3748 & 0.1747 \\
PiRD        & 0.3684 & 0.1689 \\
RSGD        & 0.3511 & 0.1498 \\
Diff        & 0.3524 & 0.1457 \\
Cond Diff   & 0.3540 & 0.1604 \\
\model      & \textbf{0.2936} & \textbf{0.1298} \\
\bottomrule
\end{tabular}
\caption{LPIPS scores on \textit{McWilliams Flow}.}
\end{table}

\begin{table}[H]
\centering
\small
\begin{tabular}{lcc}
\toprule
Model & 8$\times$ Upsampling & 4$\times$ Upsampling \\
\midrule
DiffPIR     & 0.6616 & 0.6525 \\
DPS         & 0.2148 & 0.2148 \\
ResShift    & 0.2231 & 0.2514 \\
PiRD        & 0.2262 & 0.2524 \\
RSGD        & 0.1639 & 0.2115 \\
Diff        & 0.4215 & 0.3608 \\
Cond Diff   & 0.1688 & 0.2259 \\
\model      & \textbf{0.1397} & \textbf{0.0637} \\
\bottomrule
\end{tabular}
\caption{LPIPS scores on \textit{Decaying Turbulence}.}
\end{table}

\begin{table}[H]
\centering
\small
\begin{tabular}{lcc}
\toprule
Model & 8$\times$ Upsampling & 4$\times$ Upsampling \\
\midrule
DiffPIR     & 0.5785 & 0.5702 \\
DPS         & 0.2148 & 0.2148 \\
ResShift    & 0.2559 & 0.3899 \\
PiRD        & 0.2881 & 0.2478 \\
RSGD        & 0.2148 & 0.1813 \\
Diff        & 0.2525 & 0.1494 \\
Cond Diff   & 0.1750 & 0.2362 \\
\model      & \textbf{0.1704} & \textbf{0.1339} \\
\bottomrule
\end{tabular}
\caption{LPIPS scores on \textit{Taylor Green Vortex}.}
\end{table}

The PSNR and SSIM results are reported below:

\begin{table}[H]
  \centering
\begin{tabular}{lcccc}
\toprule
\multicolumn{5}{c}{\textbf{McWilliams Flow}} \\
\midrule
\textbf{Model} & \multicolumn{2}{c}{\textbf{4x Upsampling}} & \multicolumn{2}{c}{\textbf{8x Upsampling}} \\
\cmidrule(lr){2-3} \cmidrule(lr){4-5}
& \textbf{PSNR} & \textbf{SSIM} & \textbf{PSNR} & \textbf{SSIM} \\
\midrule
DiffPIR & 20.7069 & 0.0426 & 20.2405 & 0.0381 \\
DPS & 18.5968 & 0.0257 & 18.7903 & 0.0270 \\
ResShift & 28.7484 & 0.5945 & 20.3056 & 0.2827 \\
PiRD & 25.2076 & 0.5110 & 20.2050 & 0.2784 \\
RSGD & 26.5900 & 0.6088 & 21.6394 & 0.3056 \\
Diffusion & 29.7101 & 0.6686 & 25.2200 & 0.3884 \\
Cond Diff & 29.6757 & 0.6665 & 25.1475 & 0.3823 \\
\model & \textbf{30.0540} & \textbf{0.6722} & \textbf{28.1972} & \textbf{0.4643} \\
\bottomrule
\end{tabular}
\end{table}

\begin{table}[H]
  \centering
\begin{tabular}{lcccc}
\toprule
\multicolumn{5}{c}{\textbf{Taylor Green Vortex}} \\
\midrule
\textbf{Model} & \multicolumn{2}{c}{\textbf{4x Upsampling}} & \multicolumn{2}{c}{\textbf{8x Upsampling}} \\
\cmidrule(lr){2-3} \cmidrule(lr){4-5}
& \textbf{PSNR} & \textbf{SSIM} & \textbf{PSNR} & \textbf{SSIM} \\
\midrule
DiffPIR & 30.0668 & 0.7328 & 20.2300 & 0.6633 \\
DPS & 30.3133 & 0.7261 & 23.2857 & 0.7373 \\
ResShift & 30.5201 & 0.8009 & 23.7859 & 0.6162 \\
PiRD & 28.3752 & 0.7028 & 22.1947 & 0.4981 \\
RSGD & 25.0744 & 0.7933 & 20.9531 & 0.5955 \\
Diffusion & 30.7706 & 0.8438 & 24.8698 & 0.6678 \\
Cond Diff & 23.2154 & 0.6767 & 21.3979 & 0.5959 \\
\model & \textbf{31.5277} & \textbf{0.8713} & \textbf{26.7503} & \textbf{0.7562} \\
\bottomrule
\end{tabular}
\end{table}

\begin{table}[H]
  \centering
\begin{tabular}{lcccc}
\toprule
\multicolumn{5}{c}{\textbf{Decaying Turbulence}} \\
\midrule
\textbf{Model} & \multicolumn{2}{c}{\textbf{4x Upsampling}} & \multicolumn{2}{c}{\textbf{8x Upsampling}} \\
\cmidrule(lr){2-3} \cmidrule(lr){4-5}
& \textbf{PSNR} & \textbf{SSIM} & \textbf{PSNR} & \textbf{SSIM} \\
\midrule
DiffPIR & 31.7640 & 0.5763 & 30.3585 & 0.5160 \\
DPS & 31.0463 & 0.5534 & 30.1307 & 0.5088 \\
ResShift & 29.0554 & 0.7998 & 23.6247 & 0.5798 \\
PiRD & 21.2584 & 0.7215 & 27.7594 & 0.6512 \\
RSGD & 40.5449 & 0.9259 & 35.3302 & 0.8185 \\
Diffusion & 47.2289 & 0.9571 & 40.8901 & 0.8622 \\
Cond Diff & 46.0675 & 0.9138 & 40.1089 & 0.8456 \\
\model & \textbf{48.0321} & \textbf{0.9601} & \textbf{43.0276} & \textbf{0.9295} \\
\bottomrule
\end{tabular}
\end{table}


\section{Runtime Comparison}
\label{appendix:runtime}
Table~\ref{tab:data_gen_time} presents the time required for the numerical solver to generate a single frame of low- and high-fidelity samples for each dataset. Note that adaptive time stepping with CFL condition is adopted to ensure numerical stability. The high-fidelity data $256 \times 256$ are generated using $4096 \times 4096$ and then downsampled to $256 \times 256$. All datasets are generated using pseudo spectral solver that run on GPU.

\begin{table}[H]
\centering
\begin{tabular}{|l|c|c|c|}
\hline
Dataset & $256 \times 256$ & $64 \times 64$ & $32 \times 32$ \\ \hline
\textit{Kolmogorov Flow} & 246.23 & 0.44 & 0.04 \\ \hline
\textit{McWilliams Flow} & 156.20 & 1.23 & 0.12 \\ \hline
\textit{Decaying Turbulence} & 151.70 & 1.01 & 0.06 \\ \hline
\textit{Taylor Green Vortex} & 247.03 & 1.49 & 0.22 \\ \hline
\end{tabular}
\caption{Run time of numerical solver to generate one frame for each dataset across different grid resolutions with a batch size of 10. All times are measured in seconds.}
\label{tab:data_gen_time}
\end{table}

Table~\ref{table:time} shows the time comparison across ML models for reconstructing high-fidelity data from low-fidelity inputs. 

\begin{table}[H]
\centering
\scalebox{0.9}{
\begin{tabular}{lcc}
\toprule
Method & 32$\times$32 $\rightarrow$ 256$\times$256 & 64$\times$64 $\rightarrow$ 256$\times$256 \\
\midrule
DiffPIR & 17.27 & 17.27 \\
DPS & 17.11 & 17.11 \\
ResShift & 3.96 & 3.96 \\
PiRD & 5.15 & 5.15 \\
RSGD & 6.13 & 3.14 \\
Diff & 6.10 & 3.06 \\
Cond Diff & 6.20 & 3.38 \\
\model & 6.37 & 3.27 \\
\bottomrule
\end{tabular}
}
\caption{Runtime comparison of various methods across different resolution levels using a batch size of 10, with all times measured in seconds.}
\label{table:time}
\end{table}




\section{Importance Weight Sensitivity Analysis}
\label{appendix:ses_imw}
\begin{figure*}[htbp]
    \centering
    \begin{subfigure}
        \centering
        \includegraphics[width=0.23\textwidth]{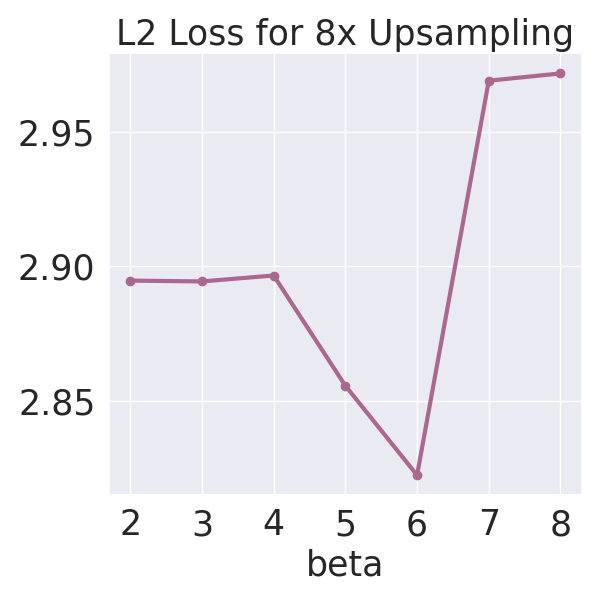}
    \end{subfigure}
    \hfill
    \begin{subfigure}
        \centering
        \includegraphics[width=0.23\textwidth]{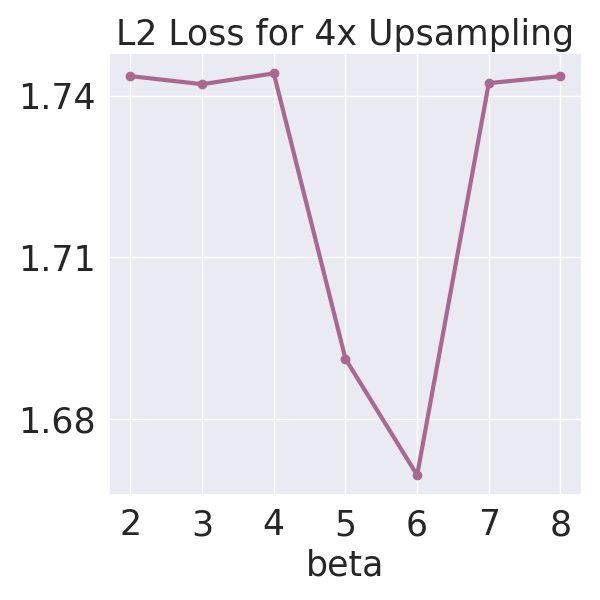}
    \end{subfigure}
    \hfill
    \begin{subfigure}
        \centering
        \includegraphics[width=0.23\textwidth]{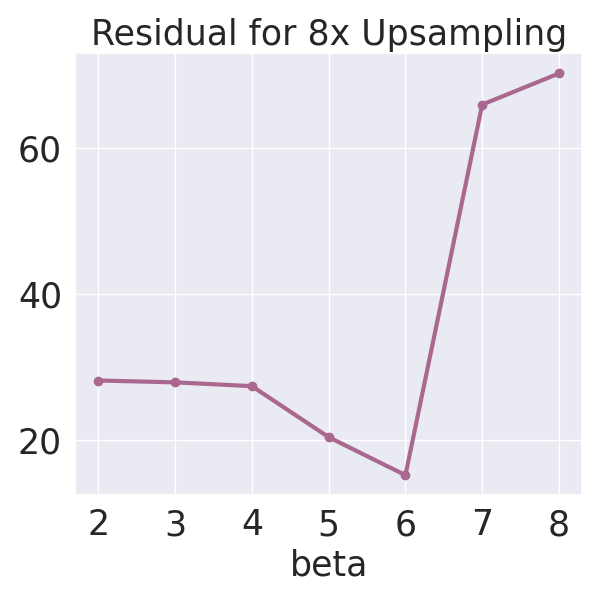}
    \end{subfigure}
    \hfill
    \begin{subfigure}
        \centering
        \includegraphics[width=0.23\textwidth]{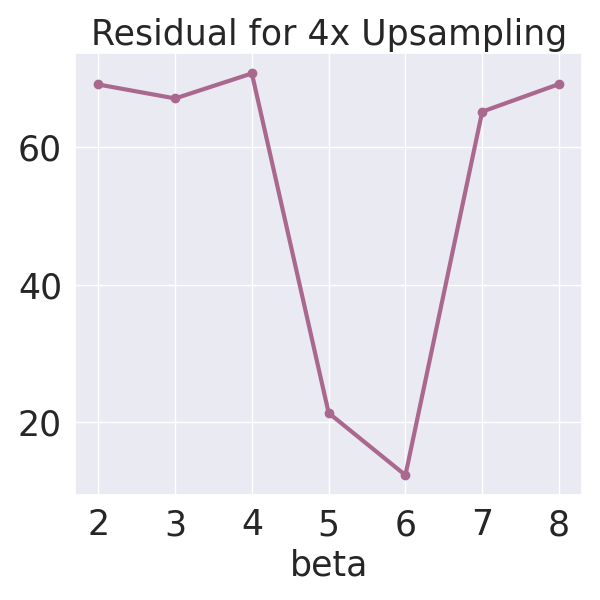}
    \end{subfigure}
    
    \vspace{0.5em}
    
    \begin{subfigure}
        \centering
        \includegraphics[width=0.23\textwidth]{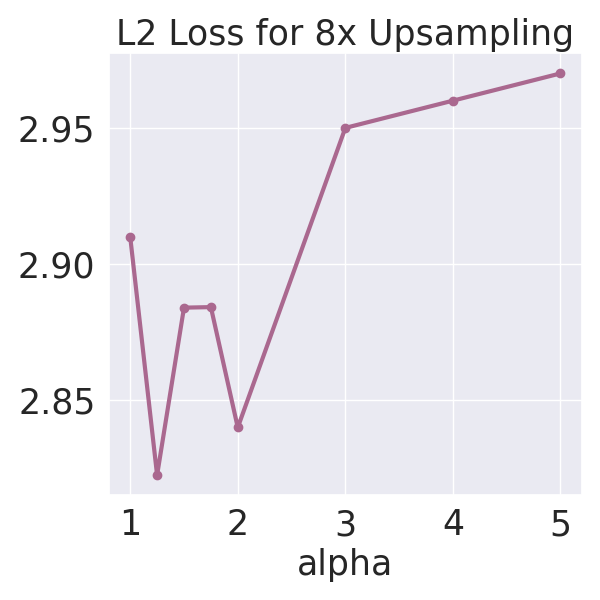}
    \end{subfigure}
    \hfill
    \begin{subfigure}
        \centering
        \includegraphics[width=0.23\textwidth]{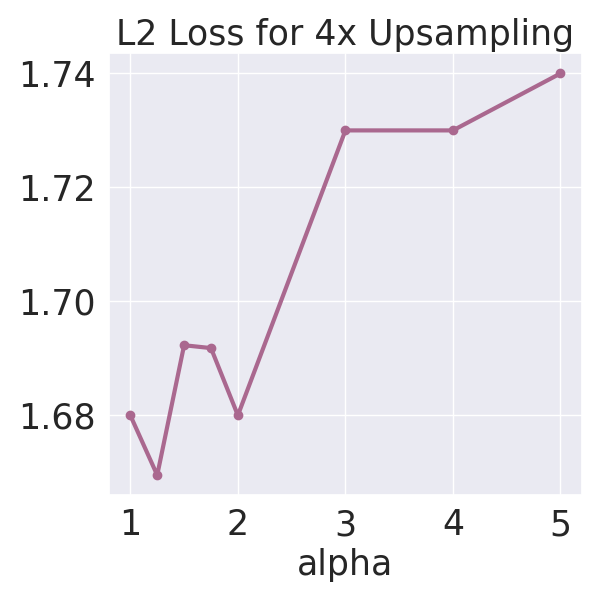}
    \end{subfigure}
    \hfill
    \begin{subfigure}
        \centering
        \includegraphics[width=0.23\textwidth]{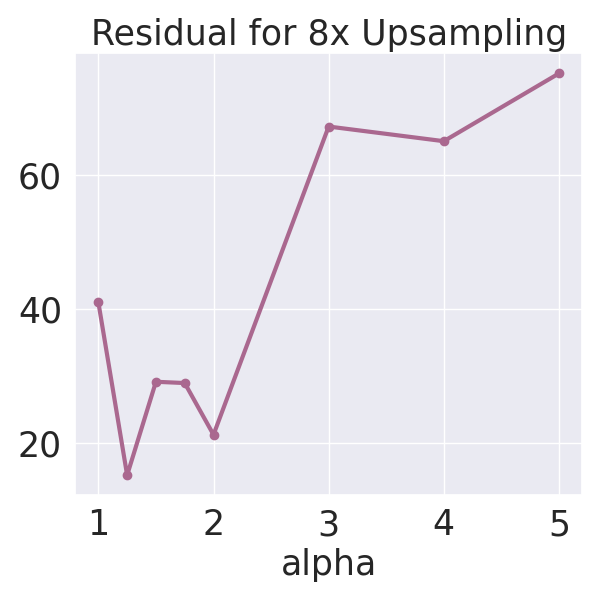}
    \end{subfigure}
    \hfill
    \begin{subfigure}
        \centering
        \includegraphics[width=0.23\textwidth]{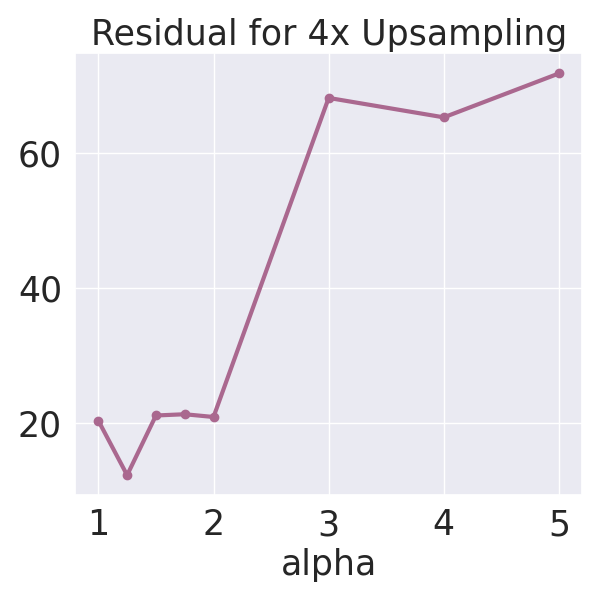}
    \end{subfigure}
    
    \vspace{0.5em}
    
    \begin{subfigure}
        \centering
        \includegraphics[width=0.23\textwidth]{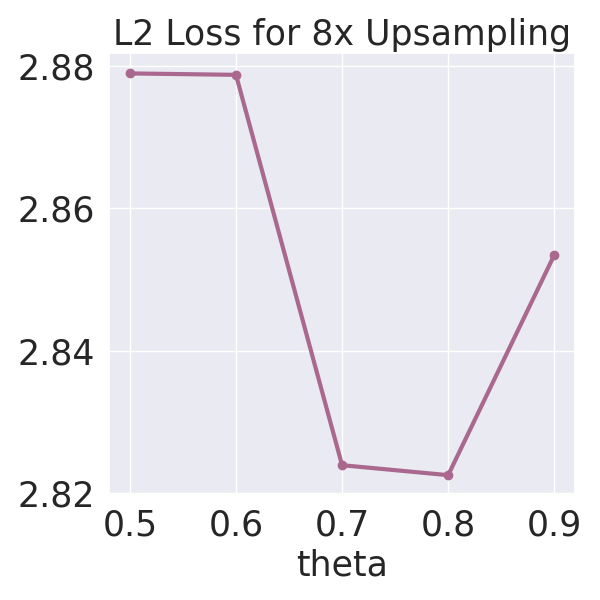}
    \end{subfigure}
    \hfill
    \begin{subfigure}
        \centering
        \includegraphics[width=0.23\textwidth]{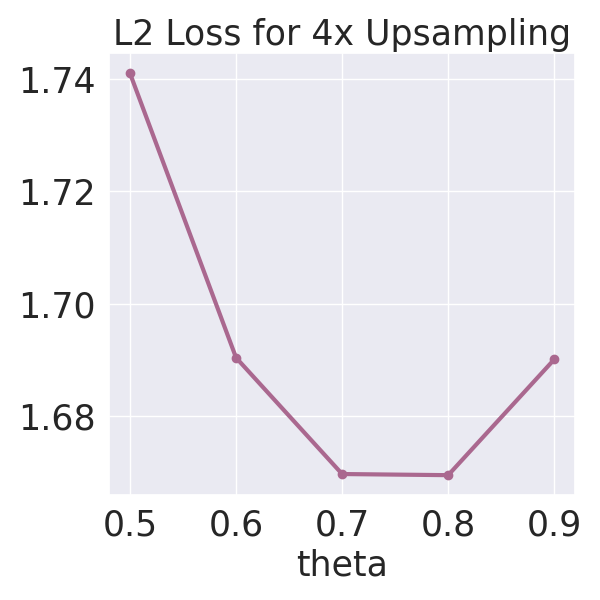}
    \end{subfigure}
    \hfill
    \begin{subfigure}
        \centering
        \includegraphics[width=0.23\textwidth]{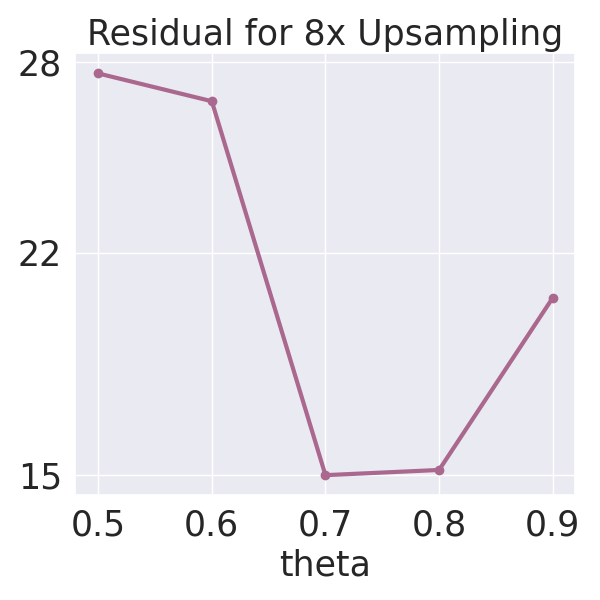}
    \end{subfigure}
    \hfill
    \begin{subfigure}
        \centering
        \includegraphics[width=0.23\textwidth]{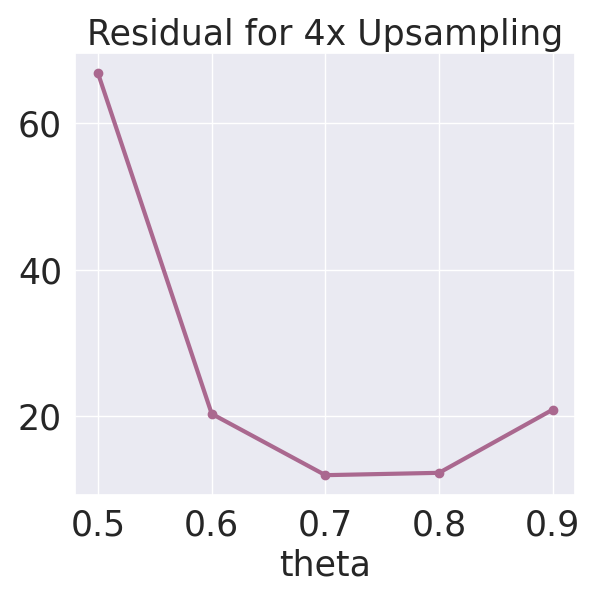}
    \end{subfigure}
    
    \caption{Sensitivity analysis of key parameters for the \textit{Importance Weight}. Experiments are conducted on the \textit{Kolmogorov Flow}. The top row presents the results for the maximum importance weight, $\beta$. The middle row displays the results for the minimum importance weight, $\alpha$, and the bottom row shows the results for the importance threshold, $\theta$. These three hyperparameters were tuned in sequence, and the optimal combination ($\beta = 6$, $\alpha = 1.25$, and $\theta = 0.8$) is selected.}
    \label{fig:ses_imw}
\end{figure*}

We present a sensitivity analysis to explore how different hyperparameter settings affect the performance of the Importance Weight Strategy. The importance weight strategy is governed by three main parameters: $\beta$ (the maximum importance weight), $\alpha$ (the minimum importance weight), and $\theta$ (the importance threshold). We adjust these parameters sequentially. The results are summarized in Figure~\ref{fig:ses_imw}.

\section{Generalization}
\label{appendix:transfer}

We observe that \model~generalizes well to low-fidelity data generated with different solver configurations. We conduct evaluations over three generalization settings: time discretization, spatial domain size, and Reynolds number as shown in Table~\ref{table:transfer_8}. We train \model~on the original \textit{Kolmogorov Flow} dataset, which has timestep $dt = 1/32$, spatial domain size $2\pi \times 2\pi$, and Reynolds number $Re = 1000$. We then directly test these new low-fidelity data on the trained model, without any additional retraining or fine-tuning. We compare the performance against models directly trained one each new configuration.

The results reveal that the pertained \model~performs comparably to trained ones directly on each new configuration. This underscores our model's strong generalization capabilities across different solver configurations. Such generalization ability is particularly important given the high-level objective is to combine \model with solvers operating on coarser grids to generate high-fidelity data faster. The generalization ability can due to the fact that both our \textit{importance weight} mechanism and \textit{residual correction} modules are training-free. They enable our model to locate fine-grained high-fidelity details and adhere to physical laws independent of training data. A similar trend is also observed in $4\times$ upsampling experiments.

\begin{table}[H]
\centering
\small
\scalebox{0.9}{
\begin{tabular}{llcc}
\toprule
\textbf{Variation} & \textbf{Model} & \textbf{L2} & \textbf{Res.} \\
\midrule
\multicolumn{4}{c}{\textbf{Time Discretization Variations}} \\
\midrule
$dt = 1/40$ & Trained on Original Data & 2.8144 & 0.0497 \\
& Trained on $dt = 1/40$ Data & 2.8136 & 0.0474 \\
\midrule
$dt = 1/50$ & Trained on Original Data & 2.7761 & 0.0253 \\
& Trained on $dt = 1/50$ Data & 2.7795 & 0.0939 \\
\midrule
\multicolumn{4}{c}{\textbf{Spatial Domain Size Variations}} \\
\midrule
$1\pi \times 1\pi$ & Trained on Original Data & 1.9045 & 2.4468 \\
& Trained on $1\pi \times 1\pi$ Data & 1.7672 & 1.1074 \\
\midrule
$1.5\pi \times 1.5\pi$ & Trained on Original Data & 2.3573 & 0.6852 \\
& Trained on $1.5\pi \times 1.5\pi$ Data & 2.3028 & 1.0282 \\
\midrule
\multicolumn{4}{c}{\textbf{Reynolds Number Variations}} \\
\midrule
$Re = 500$ & Trained on Original Data & 2.3694 & 0.2058 \\
& Trained on $Re = 500$ Data & 2.3542 & 0.1513 \\
\midrule
$Re = 2000$ & Trained on Original Data & 3.2957 & 0.2113 \\
& Trained on $Re = 2000$ Data & 3.2914 & 0.2406 \\
\bottomrule
\end{tabular}}
\caption{Generalization results on \textit{Kolmogorov Flow} dataset with $32 \times 32 \rightarrow 256 \times 256$ setting.}
\label{table:transfer_8}
\end{table}

\begin{table}[H]
\centering
\small
\caption{Generalization results on \textit{Kolmogorov Flow} dataset with $64 \times 64 \rightarrow 256 \times 256$ setting.}
\label{table:transferability_results_4x}
\scalebox{0.9}{
\begin{tabular}{llcc}
\toprule
\textbf{Variation} & \textbf{Model} & \textbf{L2} & \textbf{Res.} \\
\midrule
\multicolumn{4}{c}{\textbf{Time Discretization Variations}} \\
\midrule
$dt = 1/40$ & Trained on Original Data & 1.6782 & 0.0820 \\
& Trained on $dt = 1/40$ Data & 1.6723 & 0.0897 \\
\midrule
$dt = 1/50$ & Trained on Original Data & 1.6484 & 0.0581 \\
& Trained on $dt = 1/50$ Data & 1.6489 & 0.1567 \\
\midrule
\multicolumn{4}{c}{\textbf{Spatial Domain Size Variations}} \\
\midrule
$1\pi \times 1\pi$ & Trained on Original Data & 1.0255 & 2.4649 \\
& Trained on $1\pi \times 1\pi$ Data & 0.9443 & 1.3626 \\
\midrule
$1.5\pi \times 1.5\pi$ & Trained on Original Data & 1.3212 & 0.8261 \\
& Trained on $1.5\pi \times 1.5\pi$ Data & 1.2918 & 1.3220 \\
\midrule
\multicolumn{4}{c}{\textbf{Reynolds Number Variations}} \\
\midrule
$Re = 500$ & Trained on Original Data & 1.3112 & 0.2613 \\
& Trained on $Re = 500$ Data & 1.2998 & 0.2301 \\
\midrule
$Re = 2000$ & Trained on Original Data & 1.9647 & 0.2066 \\
& Trained on $Re = 2000$ Data & 1.9694 & 0.2035 \\
\bottomrule
\end{tabular}}
\end{table}




\section{Proof of Proposition~\ref{prop:error_bound}}
\label{appendix:proof_error_bound}
Give the Predictor-Corrector-Advancer SDE solver as follows:
\begin{itemize}
    \item Predictor $\tilde{x}_0^{t} = \text{NN}_{\theta} (x_{t}, t)$
    \item Corrector $\tilde{x}_0^{t} = \mathcal{C}(\tilde{x}_0^{t})$ if correction is applied else $\overline{x}_0^{t} = \tilde{x}_0^{t}$
    \item Advancer $x_{t-1} = x_{t} + \left[  f(x_t,t) + g^2(t) \frac{x_{t} - \tilde{x}_0^{t} - \int_{0}^{t} f(x_s, s) ds}{\int_{0}^{t} g^2(s)} \right] dt + g(t) d\overline{W}$
\end{itemize}
We can consider the SDE of interest to be
\begin{align*}
    dx = \left[  f(x,t) + g^2(t) \frac{x_{t} - \tilde{x}_0^{t} - \int_{0}^{t} f(x,s) ds}{\int_{0}^{t} g^2(s)} \right] dt + g(t) d\overline{W}
\end{align*}
Define $\overline{f}(x,t) = f(x,t) + g^2(t) \frac{x_{t} - \tilde{x}_0^{t} - \int_{0}^{t} f(x,s) ds}{\int_{0}^{t} g^2(s)}$ and $\overline{g}(t) = g(t)$.
\begin{assumption}
Let $f(x,t)$ and $g(t)$ be the drift function and diffusion coefficient of VP-SDE \citep{ho2020denoising}. Then, $f(x,t)$ is Lipschitz with respect to $x$ with Lipschitz constant $L_f$.
\end{assumption}
Then, if $\hat{x}_t$ denotes the piecewise constant solution, for $0 \leq t \leq T$,
\begin{align*}
    Z(t)
    & = \sup_{0 \leq s \leq t} \mathbb{E} \left[ |x_s - \hat{x}_s |^2 \right] \\
    & = \sup_{0 \leq s \leq t} \mathbb{E} \mid \int_{0}^{t_{n_s}}  \overline{f}(x_u, t_u) - \overline{f}(\hat{x}_u, t_u)\ du \\
    & + \int_{0}^{t_{n_s}} \overline{g} (t_u) - \overline{g} (t_u)\ dW_u \\
    & + \int_{t_{n_s}}^s \overline{f}(\hat{x}_u, t_u)\ du + \int_{t_{n_s}}^s \overline{g} (t_u)\ dW_u \mid^2 \\
    & = \sup_{0 \leq s \leq t} \mathbb{E} \mid \int_{0}^{t_{n_s}}  \overline{f}(x_u, t_u) - \overline{f}(\hat{x}_u, t_u)\ du \\
    & + \int_{t_{n_s}}^s \overline{f}(\hat{x}_u, t_u)\ du
    + \int_{t_{n_s}}^s \overline{g} (t_u)\ dW_u \mid^2 \\
\end{align*}
By Cauchy Schwarz inequality
\begin{align*}
    & \leq 3 \sup_{0 \leq s \leq t} \mathbb{E} \mid \int_{0}^{t_{n_s}}  \overline{f}(x_u, t_u) - \overline{f}(\hat{x}_u, t_u)\ du \mid^2 \\
    & + \mid \int_{t_{n_s}}^s \overline{f}(\hat{x}_u, t_u)\ du \mid^2
    + \mid \int_{t_{n_s}}^s \overline{g} (t_u)\ dW_u \mid^2
\end{align*}
By linearity of expectation, Ito’s isometry, and Cauchy Schwarz inequality,
\begin{align*}
    & \leq 3 \sup_{0 \leq s \leq t} 
    T \mathbb{E} \left[ \int_{0}^{t_{n_s}}  \mid \overline{f}(x_u, t_u) - \overline{f}(\hat{x}_u, t_u) \mid^2 du \right] \\
    & + \Delta t \mathbb{E} \left[ \mid \int_{t_{n_s}}^s \mid \overline{f}(\hat{x}_u, t_u) \mid^2 du \right]
    + \mathbb{E} \left[ \int_{t_{n_s}}^s \mid \overline{g} (t_u) \mid^2 du \right]
\end{align*}
Note that in $\overline{f}(x_t,t) = f(x_t,t) + g^2(t) \frac{x_{t} - \overline{x}_0^{t} - \int_{0}^{t} f(x,s) ds}{\int_{0}^{t} g^2(s)}$, $f(x_t,t)$ is Lipchitz with respect to $x_t$ by construction. $g^2(t)$, $\int_{0}^{t} f(x_s,s) ds$, and $\int_{0}^{t} g^2(s)$ are constant with respect to $x_t$.

\begin{assumption}
The predictor $\tilde{x}_0^{t} = \text{NN}_{\theta} (x_{t}, t)$ is Lipschitz with Lipschitz constant $L_{\text{NN}}$.
\end{assumption}

The Lipschitz assumption for the predictor is simply requiring the gradient with respect to the input $x_t$ to be bounded. This is a necessary condition for training $\text{NN}_{\theta}$.

\begin{assumption}
Let $\mathcal{R}$ be the PDE residual. Then, $\nabla \mathcal{R}(x)$ and $\nabla^2 \mathcal{R}(x)$ are bounded within the domain of interest. Further, $\nabla \mathcal{R}(x)$ is Lipschitz with respect to $x$ with Lipschitz constant $L_{\mathcal{R}}$.

\end{assumption}
The corrector $\overline{x}_0^{t} = \mathcal{C}(\tilde{x}_0^{t})$ involves $M$ steps of gradient descent using the gradient of the PDE resiudal. In the context of PDEs—particularly for the turbulent Navier–Stokes equations—this assumption may not hold due to the inherently irregular and multi-scale nature of turbulent flows. Despite this, \citet{shu2023physics} incorporate $\nabla \mathcal{R}(x)$ as an input to the denoiser, a procedure that necessitates the boundedness of both $\nabla \mathcal{R}(x)$ and $\nabla^2 \mathcal{R}(x)$ to ensure stable gradient propagation. Consequently, we adopt the assumption that $\nabla \mathcal{R}(x)$ and $\nabla^2 \mathcal{R}(x)$ are bounded within the domain of interest.

For the Adam gradient descent, $m_{t+1} = \beta_1 m_t + (1-\beta_1) \nabla \mathcal{R}(x_t)$ and $v_{t+1} = \beta_2 v_t + (1-\beta_2) (\nabla \mathcal{R}(x_t))^2$. The update rule can be written as $x_{t+1} = x_t - \eta Q(m_{t+1}, v_{t+1})$ with $Q(m, v) = \frac{m}{\sqrt{v} + \epsilon}$.
\begin{align*}
    \frac{\partial Q}{\partial \nabla \mathcal{R}} = \frac{\frac{\partial m}{\partial \nabla \mathcal{R}} (\sqrt{v} + \epsilon)  + m \frac{1}{2} v^{-1/2} \frac{\partial v}{\partial \nabla \mathcal{R}}}{(\sqrt{v} + \epsilon)^2}
\end{align*}
Using the boundedness of $\nabla \mathcal{R}$ and the Adam parameters, $Q$ is Lipschitz with respect to $\partial \nabla \mathcal{R}$ and assume the Lipschitz constant is $L_Q$. Then, $Q$ is 
Lipschitz with respect to $x$ with Lipschitz constant $L_Q L_{\mathcal{R}}$. As a result, the one step gradient descent $T(x) = x - \eta Q(x)$ has Lipschitz constant $1+\eta L_Q L_{\mathcal{R}}$. By induction, $M$ steps of gradient descet has Lipschitz constant $(1+\eta L_Q L_{\mathcal{R}})^M$.

As a result, The corrector can also be considered as Lipchitz with Lipchitz constant $L_{\overline{f}} = L_f + \frac{|\sup_{0 \leq s \leq T} g^2 (s)|}{|\sup_{0 \leq s \leq T} \int_0^s g^2(s)|} (1+(1+\eta L_Q L_{\mathcal{R}})^M L_{\text{NN}})$ if correction is applied and $\hat{L}_{\overline{f}} = L_f + \frac{|\sup_{0 \leq s \leq T} g^2 (s)|}{|\sup_{0 \leq s \leq T} \int_0^s g^2(s)|} (1+ L_{\text{NN}})$ otherwise. Note that since $(1+\eta L_Q L_{\mathcal{R}})^M > 1$, $L_{\overline{f}} > \hat{L}_{\overline{f}}$.

Let $A \cup B = [0, t_{n_s}]$ and $A \cap B = \emptyset$. Let $A$ denotes the set where correction is applied and $B$ denotes the set where correction is not applied. Then,
\begin{align*}
    & \int_{0}^{t_{n_s}}  \mid \overline{f}(x_u, t_u) - \overline{f}(\hat{x}_u, t_u) \mid^2 du \\
    & = \int_A \mid \overline{f}(x_u, t_u) - \overline{f}(\hat{x}_u, t_u) \mid^2 du
     + \int_B \mid \overline{f}(x_u, t_u) - \overline{f}(\hat{x}_u, t_u) \mid^2 du \\
     & \leq L_{\overline{f}} \int_A \mid x_u - \hat{x}_u \mid^2 du
     + \hat{L}_{\overline{f}} \int_B \mid x_u - \hat{x}_u \mid^2 du \\
\end{align*}
Define $\alpha = \frac{\int_{A} \mid x_u - \hat{x}_u \mid^2 du}{\int_{0}^{t_{n_s}} \mid x_u - \hat{x}_u \mid^2 du} \propto |A|$. Then, $1-\alpha = \frac{\int_{B} \mid x_u - \hat{x}_u \mid^2 du}{\int_{0}^{t_{n_s}} \mid x_u - \hat{x}_u \mid^2 du} \propto |B|$. We have,
\begin{align*}
    & \leq (\alpha L_{\overline{f}} +  (1-\alpha) \hat{L}_{\overline{f}}) \int_{0}^{t_{n_s}} \mid x_u - \hat{x}_u \mid^2 du
\end{align*}

Then we have
\begin{align*}
    & \leq 3 \sup_{0 \leq s \leq t} 
    (\alpha L_{\overline{f}} +  (1-\alpha) \hat{L}_{\overline{f}})^2 T \mathbb{E} \left[ \int_{0}^{t_{n_s}} \mid x_u - \hat{x}_u \mid^2 du \right] \\
    & + \Delta t \mathbb{E} \left[ \mid \int_{t_{n_s}}^s \mid \overline{f}(\hat{x}_u, t_u) \mid^2 du \right]
    + \mathbb{E} \left[ \int_{t_{n_s}}^s \mid \overline{g} (t_u) \mid^2 dW_u \right]
     \\
    & \leq 3 \sup_{0 \leq s \leq t} (
    (\alpha L_{\overline{f}} +  (1-\alpha) \hat{L}_{\overline{f}})^2 T \int_{0}^{t_{n_s}} Z(u) du \\
    & + \Delta t \mathbb{E} \left[ \mid \int_{t_{n_s}}^s \mid \overline{f}(\hat{x}_u, t_u) \mid^2 du \right]
    + C
    )
\end{align*}
where $C \Delta t = \mathbb{E} \left[ \int_{t_{n_s}}^s \mid \overline{g} (t_u) \mid^2 dW_u \right]$. The Lipchitz condition of $\overline{f}(x,t)$ implies the linear growth condition. There exists $K_f$ such that $\mid \overline{f}(x,t) \mid \leq K_f (1+|x|)$.

\begin{align*}
    & \leq 3 \sup_{0 \leq s \leq t}
    (\alpha L_{\overline{f}} +  (1-\alpha) \hat{L}_{\overline{f}})^2 T \int_{0}^{t_{n_s}} Z(u) du \\
    & + K_f^2 \Delta t \mathbb{E} \left[ \int_{t_{n_s}}^s \mid (1+|x_u|) \mid^2 du \right]
    + C \Delta t \\
    & \leq 3 \sup_{0 \leq s \leq t} 
    (\alpha L_{\overline{f}} +  (1-\alpha) \hat{L}_{\overline{f}})^2 T \int_{0}^{t_{n_s}} Z(u) du \\
    & + \Delta t \left[ K_f \left( \Delta t + \int_{t_{n_s}}^s \mathbb{E} \left[ |x_u|^2 \right] \right)
    + C \right]  \\
    & \leq 3 (
    (\alpha L_{\overline{f}} +  (1-\alpha) \hat{L}_{\overline{f}})^2 T \int_{0}^{t_{n_s}} Z(u) du \\
    & + \Delta t \left[ K_f^2 \Delta t \left( 1 + \sup_{0 \leq t \leq T} \mathbb{E} \left[ |x_t|^2 \right] \right)
    + C \right]
    ) \\
\end{align*}

As a condition for the existence of strong solution, we assume $\sup_{0 \leq t \leq T} \mathbb{E} \left[ |x_t|^2 \right] \leq D$. Then,
\begin{align*}
    & \leq 3 (
    (\alpha L_{\overline{f}} +  (1-\alpha) \hat{L}_{\overline{f}})^2 T \int_{0}^{t_{n_s}} Z(u) du \\
    & + \Delta t \left[ K_f^2 \Delta t \left( 1 + \sup_{0 \leq t \leq T} \mathbb{E} \left[ |x_t|^2 \right] \right)
    + C \right]
    ) \\
\end{align*}
By Grönwall's inequality,
\begin{align*}
    Z(T)
    & = \sup_{0 \leq s \leq T} \mathbb{E} \left[ |x_s - \hat{x}_s |^2 \right] \\
    & \leq \Delta t \left[ K_f^2 \Delta t \left( 1 + D \right)
    + C \right] \exp(3(\alpha L_{\overline{f}} +  (1-\alpha) \hat{L}_{\overline{f}})^2 T^2) \\
\end{align*}
Now, notice that
\begin{align*}
    \alpha & = \frac{\int_{A} \mid x_u - \hat{x}_u \mid^2 du}{\int_{0}^{t_{n_s}} \mid x_u - \hat{x}_u \mid^2 du} \\
    & \leq \frac{|A| \sup_{0 \leq u \leq t_{n_s}} \mid x_u - \hat{x}_u \mid^2}{t_{n_s} \inf_{0 \leq u \leq t_{n_s}} \mid x_u - \hat{x}_u \mid^2} = c_1 |A|
\end{align*}
Additionally,
\begin{align*}
    \alpha & = \frac{\int_{A} \mid x_u - \hat{x}_u \mid^2 du}{\int_{0}^{t_{n_s}} \mid x_u - \hat{x}_u \mid^2 du} \\
    & \geq \frac{|A| \inf_{0 \leq u \leq t_{n_s}} \mid x_u - \hat{x}_u \mid^2}{t_{n_s} \sup_{0 \leq u \leq t_{n_s}} \mid x_u - \hat{x}_u \mid^2} = c_2 |A|
\end{align*}
Thus, $\alpha = O(|A|)$ and $\alpha^2 = O(|A|^2)$. Next, we notice that
\begin{align*}
    L_{\overline{f}} & = \hat{L}_{\overline{f}} + \Delta L \\
    \Delta L & = \frac{|\sup_{0 \leq s \leq T} g^2 (s)|}{|\sup_{0 \leq s \leq T} \int_0^s g^2(s)|} (1+(1+\eta L_Q L_{\mathcal{R}})^M L_{\text{NN}}) + (L_f - \hat{L}_{\overline{f}})
\end{align*}
Thus,
\begin{align*}
    \Delta L = O((1+\eta L_Q L_{\mathcal{R}})^M)
\end{align*}
Plug in $\Delta L$ into $\Lambda = \alpha L_{\overline{f}} + (1-\alpha) \hat{L}_{\overline{f}}$. We get
\begin{align*}
    \Lambda & = \hat{L}_{\overline{f}} + \alpha \Delta L \\
    \Lambda^2 & = \hat{L}_{\overline{f}}^2 + 2\alpha \hat{L}_{\overline{f}} \Delta L + \alpha^2 \Delta L^2 \\
    \Lambda^2 & \leq c_3 \alpha^2 \Delta L^2 = O(|A|^2 (1 + \eta L_Q L_{\mathcal{R}})^{2M})
\end{align*}
Thus,
\begin{align*}
    Z(T)
    & = \sup_{0 \leq s \leq T} \mathbb{E} \left[ |x_s - \hat{x}_s |^2 \right] \\
    & = O ( \Delta t \exp( O( |A|^2 (1 + \eta L_Q L_{\mathcal{R}})^{2M} ))
\end{align*}

\section{Hardware Specification}
We implement all models in PyTorch. All experiments are run on servers/workstations with the following configuration:

\begin{itemize}
    \item 80 CPUs, 503G Mem, 8 x NVIDIA V100 GPUs.
    \item 48 CPUs, 220G Mem, 8 x NVIDIA TITAN Xp GPUs.
    \item 96 CPUs, 1.0T Mem, 8 x NVIDIA A100 GPUs.
    \item 64 CPUs, 1.0T Mem, 8 x NVIDIA RTX A6000 GPUs.
    \item 224 CPUs, 1.5T Mem, 8 x NVIDIA L40S GPUs.
\end{itemize}

\end{document}